\documentclass[manuscript]{acmart}

\AtBeginDocument{%
  \providecommand\BibTeX{{%
    \normalfont B\kern-0.5em{\scshape i\kern-0.25em b}\kern-0.8em\TeX}}}

\setcopyright{acmcopyright}
\copyrightyear{2024}
\acmYear{2024}
\acmDOI{XXXXXXX.XXXXXXX}

\acmConference[Conference acronym 'XX]{Make sure to enter the correct
  conference title from your rights confirmation emai}{May 11--16,
  2024}{Woodstock, NY}
\def\authnote{0} 
\def\revview{0} 
\def\revtwoview{0} 
\renewcommand{\paragraph}[1]{\vspace*{6pt}\noindent\textbf{#1.}\;}

\newcommand{\secref}[1]{Section~\ref{#1}}
\newcommand{\figref}[1]{Figure~\ref{#1}}
\newcommand{\tabref}[1]{Table~\ref{#1}}

\newcommand{\fixme}[1]{\ifnum\authnote=1{\textcolor{red}{[FIXME: #1]}}\fi}
\newcommand{\better}[1]{\ifnum\authnote=1{\textcolor{violet}{[BetterWord: #1]}}\fi}
\newcommand{\todo}[1]{\ifnum\authnote=1{\textcolor{red}{[TODO: #1]}}\fi}

\newcounter{mynote}[section]

\newlength{\saveparindent}
\newlength{\saveparskip}
\newcounter{ctr}

\newenvironment{newenum}{%
  \begin{list}{{\rm (\arabic{ctr})}\hfill}{\usecounter{ctr} \labelwidth=17pt%
      \labelsep=5pt \leftmargin=22pt \topsep=3pt%
      \setlength{\listparindent}{\saveparindent}%
      \setlength{\parsep}{\saveparskip}%
      \setlength{\itemsep}{2pt} }}{\end{list}}

\newcommand{\rev}[1]{\ifnum\revview=1{\color{blue}{#1}}\else{#1}\fi}

\newcommand{\revtwo}[1]{\ifnum\revtwoview=1{\color{blue}{#1}}\else{#1}\fi}

\newcommand{\artifactname}{\textsc{TIQA}\xspace}

  {\list{}{\leftmargin=0.1in\rightmargin=0.1in}\item[]}%
  {\endlist}

\usepackage{booktabs}
\usepackage{graphicx}
\usepackage[normalem]{ulem}
\useunder{\uline}{\ul}{}
\usepackage{rotating}
\usepackage{multirow}
\usepackage{xspace}
\usepackage{svg}

\begin{document}

\title[Trauma-Informed Qualitative Research]{Mitigating Trauma in Qualitative Research Infrastructure:\\\rev{Roles for Machine Assistance and Trauma-Informed Design}}


\author{Emily Tseng}
\affiliation{%
  \institution{Cornell University}
  \country{USA}
}

\author{Thomas Ristenpart}
\affiliation{%
  \institution{Cornell University}
  \country{USA}
}

\author{Nicola Dell}
\affiliation{%
  \institution{Cornell University}
  \country{USA}
}

\renewcommand{\shortauthors}{Tseng et al.}

\begin{abstract}
Researchers increasingly look to understand experiences of pain, harm, and marginalization via qualitative analysis.
Such work is needed to understand and address social ills, but poses risks to researchers' well-being: sifting through volumes of data on painful human experiences risks incurring traumatic exposure in the researcher.
In this paper, we explore how the principles of trauma-informed computing (TIC) can be applied to reimagine healthier tools and workflows for qualitative analysis. 
We apply TIC to create a design provocation called \artifactname, a system for qualitative coding that leverages language modeling, semantic search, and recommendation systems to measure and mitigate an analyst’s exposure to concepts they find traumatic. 
Through a formative study of \artifactname with 15 participants, we illuminate the complexities of enacting TIC in qualitative knowledge infrastructure, \rev{and potential roles for machine assistance in mitigating researchers' trauma}. 
To assist scholars in translating the high-level principles of TIC into sociotechnical system design, 
we argue for: (a) a conceptual shift from safety as exposure reduction towards safety as enablement; and (b) renewed attention to evaluating the trauma-informedness of design processes, in tandem with the outcomes of designed objects on users' well-being.
\end{abstract}




\maketitle

\section{Introduction}
\label{sec:intro}

Researchers across CSCW and HCI are increasingly concerned with digital technologies' role in individual and structural harms \cite{shelby2023sociotechnical}, like 
hate and harassment \cite{thomas2021sok}, racism \cite{erete2023method, ogbonnaya2020critical}, gender-based violence \cite{tseng2022care}, crisis and migration \cite{sabie2021migration, palen2016crisis}.
Much of this work seeks to develop rich and empathetic insight into human experience, via qualitative research methods.
Whether via primary methods (e.g., interviews, focus groups, and ethnography) or secondary methods (e.g., analyses of social media archives), qualitative inquiry is 
needed to gain fine-grained knowledge of how to mitigate these social ills, spotlight the voices of those harmed, and build empathy among the powerful.

With growing interest in these topics and methods, however, there are also concerns over the effect of doing such research on researchers.
When the labor of research involves repeatedly and deeply considering large volumes of graphic and highly personal stories of harm, researchers risk incurring a form of \textit{traumatic exposure} \cite{das2020fast, chen2022trauma}: negative psychological effects from confronting overwhelming or disturbing events, like war, violence, disaster, illness, injury, or death \cite{kleber2019trauma}.
Traumatic exposure can result in \textit{re-traumatization}, in which a person's own traumas are resurfaced, or \textit{vicarious or secondary trauma}, the psychological effects of hearing about trauma experienced by another person \cite{branson2019vicarious}.
Left unaddressed, traumatic exposure can result in traumatic stress reactions, which can harm researchers' well-being \cite{kleber2019trauma}.
An important question, then, is how to mitigate such harms while enabling important and high-quality research that grapples with societal problems. 

One potential way forward is build the tools and interfaces of qualitative data analysis in \textit{trauma-informed} ways. 
In 2022, Chen et al. \cite{chen2022trauma} proposed a framework, Trauma-Informed Computing (TIC), which urges computing scholars to acknowledge that digital technologies can cause trauma and re-traumatization, and proactively mitigate these effects.
TIC has seen rapid uptake in CSCW and HCI, in domains like social media \cite{scott2023trauma, randazzo2023if}, digital safety \cite{rabaan2023survivor, bellini2024sok}, algorithmic welfare \cite{showkat2023right, saxena2018algorithmic, bhandari2022multi}, and technologies for health, learning, and domestic care \cite{alghamdi2023co, ahmadpour2023understanding, abdulai2023trauma, bezabih2023challenges}.
\rev{TIC's suggested mitigations include a range of research, regulatory, and organizational practices, including usability heuristics, audits of machine learning (ML) systems, and workplace protections for technology companies. Less-explored, however, is how TIC can be operationalized in software design and development: how its high-level principles can guide design and engineering decisions.}

We contribute a formative study of how to deliberately enact TIC in \textit{software for mixed-initiative qualitative coding}.
A foundational practice within qualitative methods, qualitative coding involves an analyst closely reading segments of text and annotating them with their interpretations (called \textit{codes}).
\rev{In the process, analysts may repeatedly expose themselves to potentially traumatic content.}
\rev{In recent years, as the datasets available for research have grown to Internet scale, researchers have turned to \textit{mixed-initiative} software for data analysis: tools that use ML to augment human analytic practices \cite{horvitz1999principles, felix2018exploratory}.
ML assistance is now embedded in popular qualitative coding tools like NVivo, Atlas, and Marvin, via features that suggest codes, develop code ontologies, and generally 
leverage machine assistance to make coding more efficient or thorough. 
Some early research and product initiatives have also explored whether large language models (LLMs) can or should be embedded even further in qualitative research, e.g., as tools for audience adaptation, literature review, and even ideation \cite{schroeder2024large, aubin2024llms}.
However, less research has explored how mixed-initiative analysis tools might consider a different objective: helping a human coder mitigate their own traumatic exposure.

}

\rev{Motivated to address researchers' trauma, and inspired by the less-explored intersection of trauma-informed computing and mixed-initiative qualitative coding, we focus in this work on the following research questions:}
\rev{\begin{itemize}
    \item RQ1: Using qualitative coding tools as a test case, how can the high-level principles of trauma-informed computing be operationalized in the lower-level decision-making required to design and build software?
    \item RQ2: What roles could machine assistance have in mitigating the trauma possible in data-intensive research?
\end{itemize}}
\rev{Our work addresses these RQs through a formative study in two parts: a design science exercise (\secref{sec:design}) and a design provocation study conducted with 15 users (\secref{sec:study})}.
First, we applied TIC to the design of \rev{mixed-initiative} qualitative coding software, and explored the design objectives and software features it inspired.
We then built a functional prototype of the resulting design, called \artifactname (\textbf{T}rauma-\textbf{I}nformed \textbf{Q}ualitative \textbf{A}nalysis), and used it as a design provocation in an scenario-based interview study with 15 researchers from CSCW, HCI, and related disciplines whose work involves analyzing graphic, sensitive, and otherwise difficult-to-read texts.
We focused on two particularly difficult 
research areas: (1) understanding the role of technology in intimate partner violence (IPV) from firsthand accounts, and (2) analyzing hate speech and harassment on social media.
Participants used \artifactname to code a synthetic dataset representing one of the two topics, selected at the participant's discretion, and reflected on its affordances in semi-structured interviews.

Our findings illuminate the design space around  
mitigating trauma in qualitative coding
(\secref{sec:findings}, \tabref{tab:features-to-roles}).
From these findings, 
we synthesize potential desirable roles for machine assistance in mixed-initiative qualitative coding that take researchers' trauma into account (\secref{sec:dis}).
More broadly, 
we outline guidance for CSCW researchers
similarly looking to enact TIC in sociotechnical systems, by reflecting on our experiences of making tradeoffs between TIC principles (\ref{sec:dis-safety-collision}) and handling a lack of evaluative frameworks (\ref{sec:dis-eval}).
In sum, we contribute:

\begin{newenum}
    \item A design exploration of how to redesign qualitative coding to mitigate researchers' traumatic exposure, concretized in a functional software prototype called \artifactname (\secref{sec:design}).
    \item A provocation study of how artifacts like \artifactname might alter qualitative coding practices, conducted with 15 researchers who study potentially distressing and highly personal topics (\secref{sec:study}).
    \item \rev{Potential desired and undesired roles for machine assistance in trauma-informed qualitative coding workflows, both individual and collaborative (\tabref{tab:features-to-roles}).}
    \item Lessons for CSCW on how TIC's high-level principles can translate to the low-level decision-making required to build software, including a conceptual shift from safety-as-exposure-reduction towards \textbf{\textit{safety-as-enablement}} (\secref{sec:dis-safety-collision}, \ref{sec:dis-eval}).
\end{newenum}

\paragraph{Positionality}
The authors are CSCW, HCI, and computer security and privacy researchers who 
collectively have decades of experience researching technology's role in societal harms, including worker exploitation, online hate and harassment, gender-based violence, and human trafficking. 
All are trained in qualitative and quantitative research methods, and are academics at U.S. universities and research institutions.
In addition, each author has worked for 5--8 years in victim advocacy for survivors of gender-based violence. 
Each has received clinical training in trauma-informed care, and their research with trauma survivors builds on their deep and long-term partnerships with a wide range of trauma experts, including social workers and therapists. 

\section{Background and Related Work}
\label{sec:background}

\subsection{The shift towards machine assistance in qualitative research infrastructure}
\label{sec:relwork-qual}

Our work is situated in CSCW's longstanding interest in how knowledge infrastructures are changing under social, political, and technical imperatives. 
Writing in 2013, Edwards et al. \cite{edwards2013knowledge} traced how increasing adoption of personal computing and the Internet has created fundamental changes in how knowledge is constructed and disseminated: from expert, fixed, and authoritative sourcing towards the distributed and ever-changing wisdom of the crowds.
To meet the moment, Edwards et al. urged the development of long-term, large-scale, and collaborative infrastructure for qualitative research.
A foundational epistemology for rich and empathetic interpretation of human experience \cite{small2022qualitative}, qualitative research is uniquely suited for the study of sensitive and emotionally charged human experiences---like technology's role in harm.
Yet, unlike in quantitative traditions, where machine learning researchers have congregated around large-scale datasets,
Edwards et al. observed that qualitative research tools ``\textit{remain tedious, fragile, and intended for small-scale efforts}'' \cite{edwards2013knowledge}, and called for systems to support collaborative analyses at scale.

We focus on one important component of a next-generation qualitative research infrastructure: \textit{qualitative coding}, a widespread practice where one or more analysts make meaning from data by indexing and labeling it via close reading and annotation \cite{mcdonald2019reliability}.
An analyst can look across an annotated corpus to find relationships between codes that develop into themes (thematic analysis), or use codes to develop grounded theory \cite{braun2022conceptual, braun2019reflecting}.
Coding can align with deductive or inductive analyses, and often iterates through either approach. 
In the former, a coder applies an existing schema of codes, often derived from pre-existing theory, to annotate dataset in a ``top-down'' fashion (e.g., labeling a document for hate speech). In the latter, a coder freely annotates a dataset (a process called \textit{open coding}), 
and uses theory after to connect the meaning they find in the data to the literature.
Careful and theoretically solid qualitative coding is vital to empathetic and meaningful research---however, the process poses practical challenges.
Coding is laborious, requiring massive investments of time and effort, and the volume of data available for research is only growing \cite{basit2003manual}. 

Since 2013, researchers in CSCW, HCI, and the digital humanities have taken up the challenge of redesigning qualitative coding infrastructure.
\rev{Much of this work explores what \citet{horvitz1999principles} called \textit{mixed-initiative} systems: interactive tools that augment human capabilities with machine intelligence (\textit{mixing} human and machine \textit{initiative}) \cite{feuston2021putting, horvitz1999principles, ramos2020interactive, amershi2019guidelines}}. 
For example, Nelson's \cite{nelson2020computational} \textit{computational grounded theory} proposes an iterative process of studying text corpora: (1) \textit{pattern detection}, in which unsupervised ML tools help researchers inductively explore data; (2) \textit{pattern refinement}, in which researchers use deep reading of documents to understand the detected patterns; and (3) \textit{pattern confirmation}, in which ML tools are again used to assess how well a refined pattern generalizes across a corpus.
Researchers have proposed mixed-initiative tools for pattern detection and refinement
(e.g., PaTAT \cite{gebreegziabher2023patat}, Topicalizer \cite{baumer2020topicalizer}, Scholastic \cite{hong2022scholastic}), as well as  
pattern confirmation (e.g., Potato \cite{pei2022potato}), \rev{and many of the underlying techniques are now embedded in commercial tools for qualitative analysis (e.g., NVivo, Prodigy, Atlas and Marvin)}.
While these tools aim to \textit{balance} human and machine initiative, scholars caution that using ML in analysis may lead to \textit{overreliance} on machine judgment. 
\citet{baumer2017comparing} argued 
scholars should not confuse computation for sensemaking, and
\citet{jiang2021supporting} found  
computer-driven pattern detection can foreclose the ``\textit{serendipity}'' of quality qualitative work.

\rev{Debates around the appropriate use of mixed-initiative tools in qualitative coding closely mirror broader debates in ``\textit{human-machine complementarity}''---a field concerned with appropriately balancing human and machine capabilities in sociotechnical systems.
\citet{zerilli2019algorithmic} described complementarity as an optimal allocation of roles and responsibilities between human and machine agents completing a shared task.
In a complementary system, ``\textit{humans and machines have clearly defined and clearly separated roles...those subcomponents better suited for human handling are not automated, and those better suited for computer handling are not manually controlled}'' \cite{zerilli2019algorithmic}.}
Complementarity has become the subject of enormous attention with the commercial rise of ML.
\rev{As the industry's data needs outstrip what can be feasibly human-labeled, demand has exploded for datasets even partially labeled by humans---and tools for making human annotation more efficient by leveraging machine assistance.}
Many annotation tasks, like flagging user-generated social media posts for objectionable content or labeling satellite images for stop signs, \rev{bear a striking resemblance to qualitative coding (particularly of the deductive variety), and the systems designed to make annotators more efficient employ many of the same mixed-initiative approaches offered to researchers.}
\rev{In ML, for example, researchers have developed systems and algorithms for \textit{semi-supervised learning} or \textit{weak supervision}, in which a small amount of the total data is human-labeled and statistical techniques are used to extrapolate those labels to the broader set \cite{ratner2017snorkel}.
In HCI, researchers have developed frameworks like \textit{interactive machine teaching}, which uses pedagogical theory to inform how a human can teach a machine to perform a task like annotation \cite{ramos2020interactive}.
}

In our work, we extend this literature towards a less-explored area of improvement for qualitative research infrastructure: how mixed-initiative qualitative coding tools might uplift the well-being of the researchers and analysts whose labor they scaffold.
While prior work has focused on scaling up qualitative coding to meet the demands of Internet-scale data, we consider whether ML-supported workflows might support a scale-agnostic goal---mitigating traumatic exposure.
\rev{With human-machine complementarity in mind, we consider what \textit{roles} might be imagined for machine assistance in a qualitative coding process, and how to balance those roles with human agency and capability.}

\subsection{Mitigating traumatic exposure in analytic workflows}
\label{sec:relwork-trauma}

To attain rich and empathetic knowledge of human experience, qualitative research prioritizes what Small and Calarco call \textit{exposure}: a researcher's level of closeness to the subject, often measured as time spent in direct contact with the relevant informants, sites, or data  \cite{small2022qualitative}. 
From a basis of exposure, researchers can strive for \textit{empathy}---understanding and explaining participants' understanding---and \textit{palpability}---describing their experiences in details rather than abstractions.

Exposure, empathy, and palpability can lead to rich qualitative analysis---but also require analysts to work deeply with potentially sensitive, troubling, or disturbing media.
CSCW has viewed such work as \textit{traumatic exposure} \cite{steiger2021psychological}, which if left unaddressed can result in \textit{trauma}: the psychological effect of confronting overwhelming and disturbing events, like war, violence, disasters, illness and death.
Characterized by ``\textit{an extreme sense of powerlessness}'', trauma can lead a person to believe ``\textit{the obvious certainties of life have disappeared...the idea of the benevolence of the world, and the idea that other people can be trusted, are devastated}'' \cite{kleber2019trauma}.
Trauma is widespread: it is estimated that over 70\% of the worldwide population has experienced at least one traumatic event in their lifetime \cite{kleber2019trauma}.
For some, trauma can resolve to traumatic stress reactions that harm well-being.

Traumatic exposure can be a natural consequence of quality qualitative research, where a researcher's goal is to understand human experiences of pain and harm.
In some cases, a researcher may encounter a topic that recalls a traumatic experience they have personally had, and experience \textit{re-traumatization}---e.g., someone grieving the recent death of a family member may find it hard to read about family.
In other cases, they may experience \textit{secondary} or \textit{vicarious} trauma, or the cumulative effect of bearing witness to another person's distress \cite{branson2019vicarious} (also called \textit{compassion fatigue}). 
In the health services, where care workers like doctors, nurses and counselors assist people in distress or pain, vicarious trauma has been linked to high rates of burnout~\cite{adler2022burnout}, in which a worker is emotionally and physically exhausted by their occupational stress \cite{maslach1998multidimensional}.
\rev{Journalists have also noted the secondary stress incurred by reporters and editors whose jobs require regularly witnessing traumatic events, and developed training courses educating news professionals on traumatic stress \cite{thompson2021dart}.}

The effects of traumatic exposure have been of great concern in the occupational contexts of commercial content moderation and sociotechnical research. 
While there are meaningful differences in these labor conditions, these contexts share a core workflow: an analyst collects data that may or may not include content they find traumatic, and must deeply study the data as part of their job. 
Steiger et al. \cite{steiger2021psychological} observed the work of commercial content moderators often involves repeated exposure to traumatic content, in a ``\textit{ghost work}''-like occupational setting where call center conditions create feelings of disempowerment that amplify stress.
Researchers who study experiences of hate, harassment, mis- and disinformation, and marginalization have noted similar concerns over the effects of their jobs on their psyches (cf. \cite{tseng2020tools}, Starbird quoted in \cite{steiger2021psychological}).
Traumatic exposure in sociotechnical research has become so salient a concern that it motivated a CHI 2022 workshop on researcher well-being \cite{feuston2022researcher}.

Interventions for traumatic exposure include adjustments to both the occupational context of an analyst's work and the tools and interfaces they are provided to work with.
Occupational adjustments include worker well-being interventions, like structured debriefs and support from peers or trained professional counselors, or more structural interventions like reducing the volume of content a worker is expected to consume \cite{steiger2021psychological}.
In the space of tools and interfaces, scholars have suggested applying ML to measure the traumatic quality of a piece of content, as a precursor to limiting the amount of exposure an analyst must incur.
Motivated by the need for content moderators to analyze user-generated images, scholars in computer vision have investigated how grayscaling, obscuring, blurring, and otherwise editing the visual representation of an image might support analysts' well-being \cite{das2020fast, karunakaran2019testing}.
Notably, equivalent techniques are less-investigated in text data. Writing for an audience of natural language processing (NLP) researchers, Kirk et al. \cite{kirk2022handling} suggested similar visual obscuration techniques for potentially harmful text--- e.g., replacing a slur with a placeholder, or adding an overlay that says ``ABUSIVE'' in red text over it---but cautioned such techniques might ``\textit{constrain the actual work of annotation}''.

In this work, we build on these lines of scholarship to examine how the workflows and interfaces of qualitative coding can be remade to track and limit an analyst's exposure to topics they personally find traumatic.
We focus on text analysis, where flexible, personalizable, and semantically grounded filtering mechanisms are less-explored.

\subsection{Trauma-informed computing (TIC)}
\label{sec:relwork-tic}


As guidance for how to redesign qualitative coding to mitigate an analyst's traumatic exposure, we look to the recent literature in HCI and design on \textit{trauma-informed computing} (TIC). 
Writing at CHI in 2022, Chen et al \cite{chen2022trauma} proposed TIC as an overall orientation computing research, design, and development through which technologists can account for the impact of trauma in technology experience.
TIC emphasizes accounting for trauma in not only the interfaces and algorithms that structure individual interactions with computational systems, but also at the sociotechnical level, in how computing structures organizational relations and social worlds. 

To Chen et al. \cite{chen2022trauma}, a trauma-informed computing project is an ongoing commitment to first \textit{acknowledging} the impact of trauma, then \textit{recognizing} that digital technologies play a role in its persistence, and finally \textit{actively seeking out} how to avoid trauma and re-traumatization.
Drawing on the six principles for trauma-informed care outlined by the U.S. Substance Abuse and Mental Health Services Administration (SAMHSA), Chen et al. outline six principles that computing professionals can follow towards becoming trauma-informed: \textit{safety}, \textit{trust}, \textit{enablement}, \textit{peer support},  \textit{collaboration}, and \textit{intersectionality}.
A growing number of scholars have considered TIC in their efforts to address problems as diverse as digital safety \cite{zheng2024s, razi2024toward, randazzo2023if, bellini2024sok}, social services \cite{showkat2023right, saxena2018algorithmic, bhandari2022multi}, and technologies for health, learning, and domestic care \cite{alghamdi2023co, ahmadpour2023understanding, abdulai2023trauma, bezabih2023challenges}.
Notably, Scott et al. \cite{scott2023trauma} extend Chen et al. \cite{chen2022trauma} by elaborating a list of types of traumas (e.g., individual, developmental, and collective traumas) and honing in on how they might manifest in social media, specifically.
Scott et al. \cite{scott2023trauma} then adapt the Missouri Model of organizational change to consider how a social media platform might benchmark its level of trauma-awareness.
\rev{In the CSCW and design literature, scholars including \citet{randazzo2023trauma} and \citet{eggleston2024trauma} have further urged the incorporation of trauma-informed principles into such tools as usability indicators and user experience heuristics. \citet{zheng2024s} conduct a retrospective analysis of the design process around a potentially traumatic app---a data donation platform for research on sexual assault---to synthesize how designers can apply trauma-informed approaches in design goals, activities, and choices.
\revtwo{The notion of trauma-informedness has also been taken up in sociotechnical scholarship more broadly: Ramjit et al. \cite{ramjit2024navigating} applied trauma-informed care principles to the design of computer security interventions, by analyzing where in a standard intervention protocol traumatic stress reactions might occur, and redesigning a protocol accordingly.}}

Of particular interest in this growing literature is the application of TIC to scientific infrastructure. 
Both Chen et al. \cite{chen2022trauma} and Scott et al. \cite{scott2023trauma} specifically call for attention to how content moderators can manage their traumatic exposure, and how research can avoid re-traumatization.
Razi et al. \cite{razi2024toward} outline how to adapt trauma-informed practices for HCI research with youth, including how college-aged research assistants might react to analyses of safety challenges close to their experiences. 
Zheng et al. \cite{zheng2024s} worked specifically to design an app for data donation for sexual assault survivors, to enable research into their needs and experiences.
To this body of work, we contribute an enactment of TIC in the domain of qualitative coding software, where to the best of our knowledge the concept has been less-explored.

\section{Design Inquiry}
\label{sec:design}

\begin{figure}[h]
    \centering
    \includegraphics[width=\columnwidth]{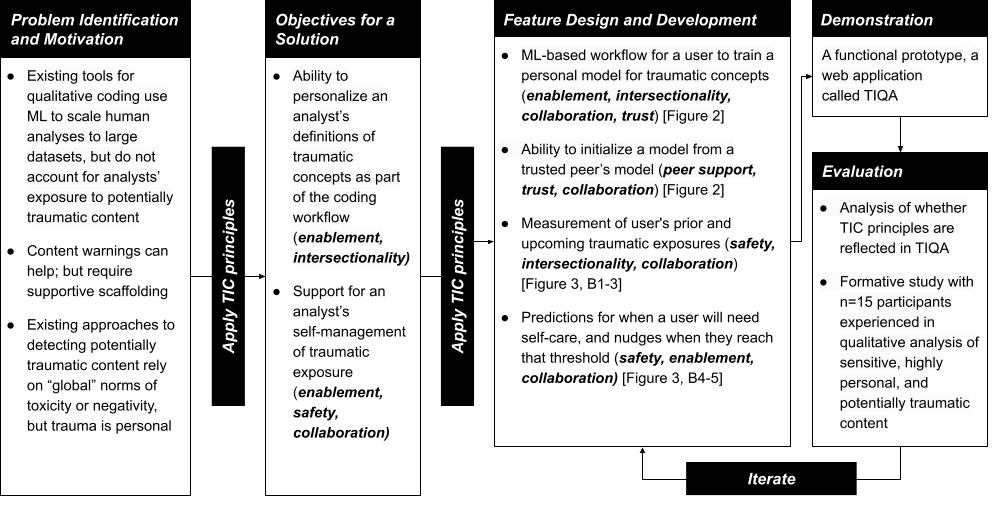}
    \caption{Design science research process, following \cite{peffers2007design} as demonstrated in \cite{reinecke2013knowing}.}
    \label{fig:design-process}
\end{figure}

\rev{We began with a design inquiry following a design science research approach drawn from \citet{zimmerman2014research}, \citet{peffers2007design}, and \citet{zheng2024s}.
As depicted in \figref{fig:design-process}, this approach consists of five steps: (1) scoping the problem at hand; (2) deciding on objectives for a solution; (3) developing specific features that realize those objectives; (4) implementing these features into a working demonstration; and (5) using the demonstration to evaluate the design.
We describe steps 1-4 in this section as our design inquiry. Step 5 is described in the following section via our formative study.
Throughout our process, we \textit{prospectively} applied \citet{chen2022trauma}'s six principles of trauma-informed computing to inspire our design decisions.
Thus we built on \citet{zheng2024s}'s \textit{retrospective} analysis of how their design process manifested trauma-informed principles.}
\revtwo{By examining prospective design decisions in this way, we addressed RQ1: how TIC could be operationalized in the lower-level decision-making required to build software.}

\subsection{\rev{Mapping problems to objectives}}

Mitigating traumatic exposure in qualitative coding is a broad design space that could take many forms. \rev{As reviewed in \secref{sec:relwork-trauma}, existing potential mitigations include both social processes (e.g., workplace support structures) and technical interventions (e.g., image obscuration techniques), as well as combinations thereof.
For this work, we decided to explore on a technical intervention, a mixed-initiative system for text analysis, because (a) such systems are recently gaining traction, with the advent of large language models (LLMs); and (b) there exists a literature gap in how technical interventions can be developed in ways that honor prosocial objectives like reducing trauma.} 

To seed our design process, we scoped the problem by drawing on both the literature and our own experiences as researchers who use qualitative coding tools in the study of potentially traumatic subjects.
\rev{Whether ML-assisted or not, the qualitative coding tools we are aware of (\secref{sec:relwork-qual}) revolve around the same core workflow:}
An analyst begins by loading one document out of a corpus and performing \textit{annotation}, the task of defining codes attached to particular segments of text.
As they proceed through the corpus, the analyst must make decisions about which documents to read and in what order; how much time to allot for their annotation work; and how they should manage the traumatic reactions they might have to the content and the work (\secref{sec:relwork-trauma}).
\rev{As surveyed in \ref{sec:relwork-qual}, existing systems employ ML assistance to augment an analyst's process (e.g., by suggesting codes or thematic ontologies); however, to our knowledge, no system uses ML assistance to mitigate the potentially traumatic effects of researchers' work.}

\rev{Taking TIC into account, we focused our design process on two main subproblems within this proposition.}
First, to minimize the effect of reading traumatic text, a system must first derive a definition of what text is traumatic.
Existing approaches to managing harmful content have leveraged block lists or global models of toxicity, e.g., the Perspective API.\footnote{https://www.perspectiveapi.com/} 
While these approaches are a good start to understanding norms that may be common in certain societies, TIC's principle of \textit{intersectionality} reminds us that our experiences of the world are levied through our specific matrices of structural oppression \cite{collins2020intersectionality}.
What reads as traumatic to one person may not read as traumatic to the next \cite{goyal2022your}.
\rev{Hence we arrived at \textit{design objective \#1: enabling an analyst to define which topics they personally might find traumatic \textit{within} their annotation workflow}.}
\revtwo{In terms of RQ1, how to operationalize TIC, this design decision reflects selecting one of its principles---in this case, intersectionality---and centering it in the system.}

\rev{Downstream of detecting which text might be potentially traumatic, a system must also decide what to do when it has identified such text.}
One approach might be to automatically obscure it from the analyst, e.g., via filters that replace sensitive text with a placeholder.
Such automatic filtering has been widely used across film, media, and classroom syllabi, and has gone by many names in the popular and academic literature, including \textit{content warnings}, \textit{trigger warnings}, or \textit{content notes}. Despite their widespread use, academic research is divided on content warnings' efficacy in reducing trauma. Some studies show they can be supportive in reducing secondary or vicarious trauma, while other studies show they have no effect, or can even be disempowering or re-traumatizing \cite{scott2023trauma}.
As a way forward through what is a yet-unsettled debate, scholars have suggested content warnings should not be employed in isolation, but rather as part of a fuller and more holistic trauma-informed approach: e.g., an individual should not be left to process a content warning alone, but rather provided with social support and self-management tools to help them through their reactions to the material \cite{bryce2023pulling}---\revtwo{echoing Chen et al.'s principles of \textit{peer support}, \textit{collaboration}, and \textit{enablement}.}
\rev{Hence we arrived at \textit{design objective \#2: supporting self-management of analysts' reactions to their own traumatic exposure.}}
\revtwo{With respect to RQ1, how to operationalize TIC, this design decision again reflects emphasizing one or more of its principles---in this case, peer support, collaboration, and enablement.}


\subsection{\rev{Mapping objectives to features}}

\rev{Having identified two objectives for our design, we mapped each to specific features of a future system.
Our resulting concept is \artifactname, a system for \textbf{T}rauma-\textbf{I}nformed \textbf{Q}ualitative \textbf{A}nalysis.}
\rev{In this section, we detail the features that arose from each design objective, and how our decisions around each feature enact the principles of TIC.}
We refer to the architectural diagrams of \artifactname in Figures~\ref{fig:tiqa_A} to \ref{fig:tiqa_B}, and describe the role of specific system modules as summarized in \tabref{tab:tiqa-modules}.
Throughout, we use italics to emphasize TIC principles.

\subsubsection{Design Objective \#1: Personalizable definitions of traumatic concepts}
\label{sec:system-personalize}

\begin{figure}[t]
    \centering
    \includegraphics[width=\columnwidth]{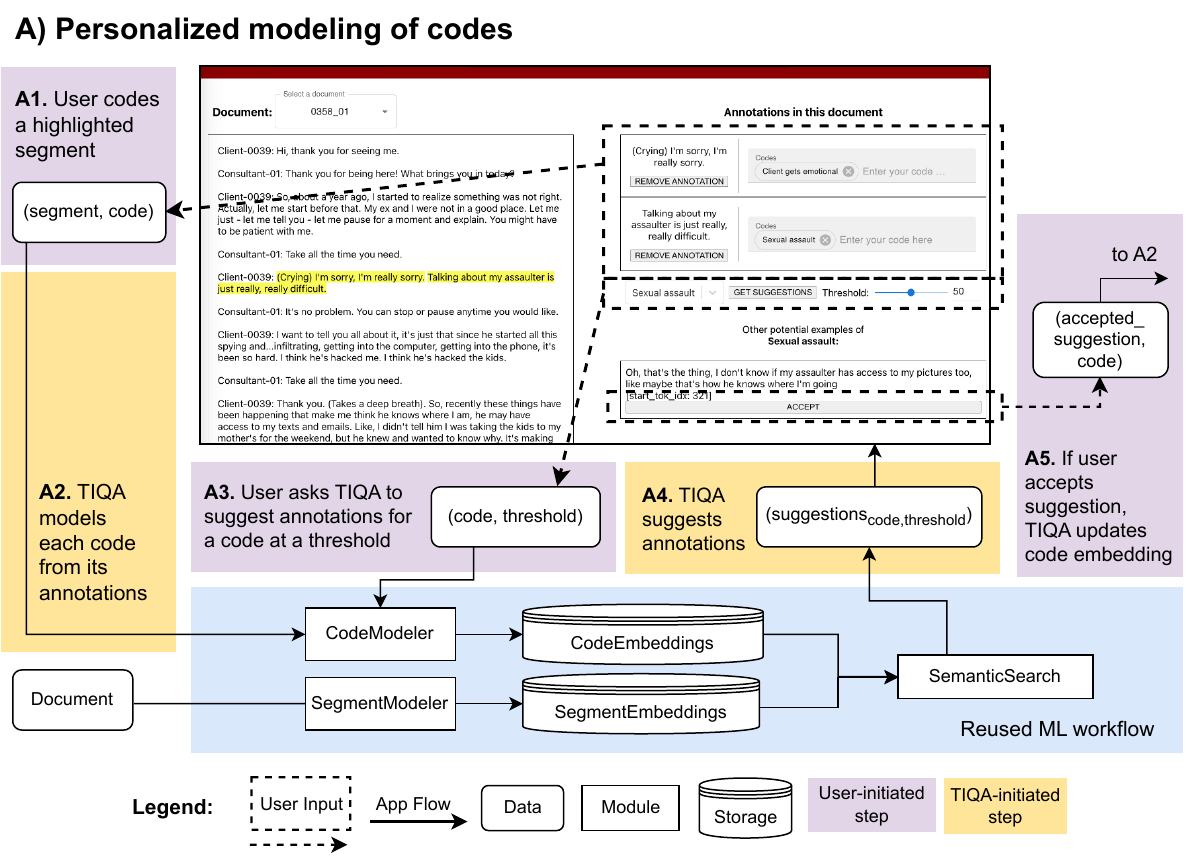}
    \caption{Architectural diagram of how \artifactname enables users to develop personalized models of their codes using an underlying ML workflow (shaded blue box). 
    See \secref{sec:system-personalize} for a step-by-step walkthrough, and \tabref{tab:tiqa-modules} for detail on each module.
    Steps are labeled A1-5, corresponding with their label in the \secref{sec:system-personalize} walkthrough.
    }
    \label{fig:tiqa_A}
\end{figure}

To \textit{enable} analysts to define for themselves what concepts they would like to be warned about, 
we sought to move away from a ``one-size-fits-all'' approach to sensemaking towards an appreciation for an individual's situated perspective (cf. feminist standpoint epistemology \cite{haraway2016situated,bardzell2010feminist}).
ML offers a possible starting point for personalized sensemaking, given the ability to fine-tune pre-trained models for specific contextual objectives. 
Many mixed-initiative qualitative coding tools (\secref{sec:relwork-qual}) already allow an analyst to build their own models for a given code.
ML-based approaches also have the benefit of capturing semantic subtleties that may be missed by keyword-based approaches like block lists. 

\begin{figure}
    \centering
    \includegraphics[width=\columnwidth]{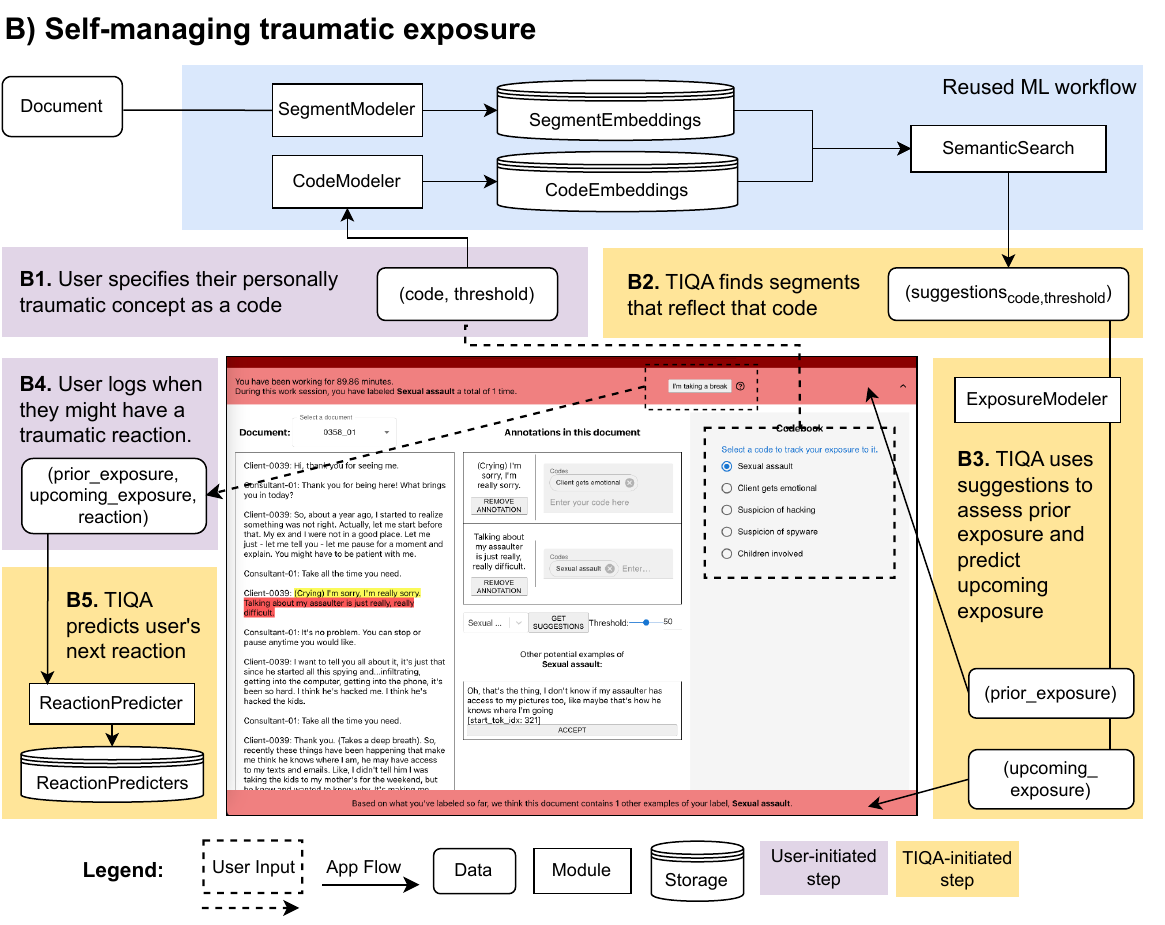}
    \caption{Architectural diagrams of how \artifactname enables users to self-manage their traumatic exposure using the same underlying ML workflow as in part A (blue box). 
    See \secref{sec:system-trauma} for a step-by-step walkthrough, and \tabref{tab:tiqa-modules} for detail on each module.
    Steps are labeled B1-5, corresponding with their label in the \secref{sec:system-trauma} walkthrough.
    }
    \label{fig:tiqa_B}
\end{figure}

\artifactname realizes this design objective via a core workflow for mixed-initiative annotation (\figref{fig:tiqa_A}), which can be applied to define traumatic concepts (\figref{fig:tiqa_B}).
A user annotates a document by highlighting a segment of the text and attaching one or more free-text codes to it (\textbf{A1}).
Codes can be developed in-situ, or pulled from an existing codebook via a dropdown menu. 
\artifactname uses the user's annotations to develop a model for the code's semantics
---a \textit{code embedding}---via the CodeModeler module 
(\textbf{A2}).
Then, \artifactname provides the user with the option to request \textit{suggestions} for further annotations for that code within the document---segments that match their code above a certain similarity threshold (\textbf{A3}).
\artifactname locates those segments via the \textit{SemanticSearch} module (\textbf{A4}), and displays them to the user with the option to accept each suggestion.
Accepted suggestions are then used to update the model, by incorporating the suggested segment as another annotation for that code (\textbf{A5}).

This core workflow for personalized codes supports the user's everyday work of annotation---and can also be adapted towards tracking and self-managing their traumatic exposure (\figref{fig:tiqa_B}).
The user can designate certain codes as traumatic, and use the same annotation and suggestion affordances to refine their model for that traumatic concept as they work through a corpus of documents (\textbf{B1, B2}).
\artifactname enacts the TIC principles of \textit{enablement} and \textit{intersectionality}, by giving a user control over defining what is traumatic to them.
Alternatively, if they do not wish to build their own model of traumatic concepts, they may import a code embedding from a trusted peer (akin to the filter import features of content moderation tools like FilterBuddy \cite{Jhaver:2022FilterBuddy})---e.g., a user sensitive to discussion of sexual assault might import a trusted peer's model for that concept. 
In this, we sought to enact TIC's principle of \textit{peer support}, which encourages computing systems to enable connections between users that can be helpful for healing.
\revtwo{Thus, with respect to RQ1, we again chose specific TIC principles to guide the selection of specific features, this time at a lower level of granularity.}

\subsubsection{Design Objective \#2: Supporting an analyst's self-management of traumatic exposure}
\label{sec:system-trauma}

We explored the possibility of supportive scaffolding for content warnings by applying TIC's principle of \textit{enablement}, which emphasizes equipping people to control their own technology experiences.
Towards enabling analysts to self-manage their traumatic exposures, we looked to personal informatics, which has explored techniques for helping people track personally relevant information and use it to guide supportive actions~\cite{li2010stage, epstein2020mapping}.
These techniques typically involve collecting longitudinal data on a user's circumstances and behaviors (e.g., what they eat, how often they exercise) and what their desired outcome might be (e.g., improving their physical health).
That information can then be used in a self-learning recommender system that predicts whether a given behavior might help them achieve their desired outcome, and surfaces those predictions to the user for self-reflection. 
Some systems leverage these predictions alongside behavioral science techniques like gamification and nudging, to support actions towards their goals \cite{okeke2018good}.

We adapted these techniques for traumatic exposure by leveraging the personalized traumatic concept models from Objective \#1 into a system that equips an analyst to self-reflect on when their exposures might cause traumatic reactions, and nudges them to engage in self-care.
When a user opts in, \artifactname displays two panes at the top and bottom of the page. 
The top pane shows the user's \textit{prior exposure} to that traumatic concept, defined as the number of times they have applied that code in the document (\textbf{B3}, top).
This pane also shows how long the user has spent coding, measured as the length of time from when they opened the current browser window to the time they clicked {\it "track exposure"}.
The bottom pane displays a user's \textit{upcoming exposure}: how many times SemanticSearch infers the exposure code will appear in the rest of the document (\textbf{B3}, bottom).
Together, these features enact TIC's principle of \textit{safety}: they provide a user with information on their prior and upcoming exposures to traumatic stress.

When a user reaches a point in their work where they feel a traumatic reaction might result from their exposures, they may hit the button that says ``\textit{I'm taking a break}'' (\textbf{B4}).
This provides explicit feedback to the system that given the present set of prior exposures, 
the user sensed a traumatic reaction might soon occur, and elected to take some space and time away from the work.
The next time a similar set of exposure parameters is reached, the system can then provide a nudge to the user encouraging them to again take a break (\textbf{B5}).
Thus, our design for \artifactname firmly retains user control or \textit{enablement} in pursuit of \textit{safety}---they may opt in or out of the tracking and nudges, and nudges do not alter their experience, but rather provide notifications they may dismiss at will.
As a self-learning system, this feature also espouses \textit{intersectionality}, by enabling a user to develop their own model for traumatic reactions.
\revtwo{Here, again, we operationalized TIC (RQ1) by choosing specific TIC principles to inspire specific lower-level features. We reflect further on this mode of operationalization in \secref{sec:dis-safety-collision}.}

\subsection{Implementation}
\label{sec:design-implementation}
To investigate the \artifactname concept more deeply, we implemented a functional prototype for use in a formative study with potential users (\secref{sec:study}).
\rev{Given the formative nature of our research questions, we opted to evaluate via a provocation study, in which an artifact is presented in prototype format (functional but not yet fully polished), participants are asked to interact with it, and researchers seek to interpret their reactions to its potential affordances.
Provocation studies are used in CSCW and HCI to seed participants' imaginations of possible sociotechnical futures, without prescriptively stifling their feedback \cite{tseng2020we}.} 

We used a functional prototype of \artifactname, rather than relying on storyboards or elicitation techniques, to prompt participants' reflections on the ``\textit{fuzzy, open-ended}'' interactions that can arise due to the unpredictable outputs characteristic of ML systems \cite{yang2020re}.
\rev{We also opted to create the prototype as a custom and locally hosted web application, rather than adapt existing commercial platforms, to focus our participants' attention on the specific features we wanted to study in a unified way. 
By building our own prototype, we were able to present all of our desired features in one interface, and avoided the attention cost of switching between existing platforms, or locating within a larger platform a specific feature of interest.}

Our implementation is a browser-based tool consisting of a frontend web application built with React (a JavaScript framework) and a backend server and API built with Django (a Python framework) over a SQLite database. 
The tool is self-contained, relying only on a cache of the open-source sentence embedding of a user's choice, and does not require transmitting any user input to an externally hosted language model.
This decision was to mitigate the potential privacy harms of transmitting user data to an externally hosted language model (see Ethics in \secref{sec:study}).

\begin{table}[]
\centering
\resizebox{\columnwidth}{!}{%
\begin{tabular}{@{}lll@{}}
\toprule
\textbf{Module} & \textbf{Role in TIQA} & \textbf{Implementation Approach} \\ \midrule
CodeModeler & \begin{tabular}[c]{@{}l@{}}Represents a user's personal\\ concept of a code, and stores\\ it in CodeEmbeddings. Refines\\ representation based on user \\ feedback (e.g., new examples).\end{tabular} & \begin{tabular}[c]{@{}l@{}}A code is the average embedding of the segments to which it is annotated.\\ User may add the semantic meaning of the code itself, or use only its \\ annotated segments.\end{tabular} \\ \midrule
SegmentModeler & \begin{tabular}[c]{@{}l@{}}Splits a document into segments.\\ Represents each segment as a \\ vector, which is then stored in \\ SegmentEmbeddings.\end{tabular} & \begin{tabular}[c]{@{}l@{}}A segment is an n-gram of the text, delimited by sentence boundaries.\\ A segment embedding is generated via SentenceTransformer.\end{tabular} \\ \midrule
SemanticSearch & \begin{tabular}[c]{@{}l@{}}Locates suggestions: document \\ segments which are conceptually \\ similar to a given code.\end{tabular} & \begin{tabular}[c]{@{}l@{}}Given a code, a document, and a similarity threshold (a number 0-100), \\ first locate the code's embedding and the embeddings of all segments \\ within the document. Return the segments where the cosine similarity \\ between the segment and the code exceeds the threshold.\end{tabular} \\ \midrule
ExposureModeler & \begin{tabular}[c]{@{}l@{}}Measures a user's prior and \\ upcoming exposures to a given \\ concept (the exposure code).\end{tabular} & \begin{tabular}[c]{@{}l@{}}Given an exposure code, a similarity threshold, and a document, use \\ SemanticSearch to measure how many occurrences of the exposure\\ code the user has already read, and how many are in upcoming segments.\end{tabular} \\ \midrule
ReactionPredicter & \begin{tabular}[c]{@{}l@{}}Predicts whether a user will have \\ a traumatic reaction given a set of\\ exposures. If yes, issues nudge \\ encouraging them to self-care.\end{tabular} & \begin{tabular}[c]{@{}l@{}}Given a user's explicit feedback about when they have neared a traumatic\\ reaction, train a three-feature prediction model that infers when they will \\ next near a traumatic reaction from (i) their prior exposure; (ii) their time \\ working; and (iii) their upcoming exposure.*\end{tabular} \\ \bottomrule
\end{tabular}%
}
\caption{Modules of TIQA, their roles within the overall system design, and our approach to their implementation in the present tool.
Each module can be refined by having the user contribute more and more training data, and using a more sophisticated implementation approach as the data available grows larger.
*For the purposes of our formative study, we did not implement ReactionPredicter in full, since training it would require long-term data from a user on their traumatic reactions.}
\label{tab:tiqa-modules}
\end{table}

\artifactname's various ML-based modules can be customized to a developer's desired modeling approach.
\tabref{tab:tiqa-modules} summarizes the algorithms we chose for our functional prototype.
In the core ML workflow code personalization (blue boxes in Figures \ref{fig:tiqa_A} and \ref{fig:tiqa_B}), we used sentence embeddings \cite{subramanian2018learning} to build vector representations of codes and segments.
Specifically, we used the SentenceTransformers framework\footnote{https://www.sbert.net/} from Reimers et al. \cite{reimers-2019-sentence-bert} with `\textit{all-MiniLM-L6-v2}', an open-source sentence model that maps inputs below 256 characters into a 384-dimensional vector space.\footnote{https://huggingface.co/sentence-transformers/all-MiniLM-L6-v2}
CodeModeler (Figure \ref{fig:tiqa_A})
builds models for each code from the \textit{average} of the embeddings of its constituent annotations.
The user can elect to initialize a code's embedding with an embedding of the code's text label, thereby including its semantics.
SegmentModeler (Figure \ref{fig:tiqa_A}) represents each segment as its direct embedding.
SemanticSearch (Figure \ref{fig:tiqa_A}) was implemented by measuring the cosine similarity between a code embedding and the embedding of a candidate document segment, and returning only those segments whose cosine similarity exceed a user-provided threshold.
To make the search space of candidate segments tractable, SegmentModeler processes each document into n-grams delimited by sentence boundaries.

In the traumatic exposure tracking features (Part \textbf{B}, \figref{fig:tiqa_A}), ExposureModeler measures prior exposure as (i) the number of times a user has used the exposure code in the present document and (ii) the total time working in the present session, measured as the browser session length.
If the user elects to take a break, the time working counter pauses its measurement until the user returns.
Upcoming exposure is measured as (iii) the number of times the exposure code is predicted to appear in the part of the document after the last annotation.
For the purposes of our study, we did not implement a full ReactionPredicter module---it was not necessary for our formative study, which could not meaningfully evaluate customized predictions in short sessions with participants (as detailed in the next section).
For a longitudinal study, ReactionPredicter may be implemented as a simple binary classifier using (i), (ii), and (iii) as features to predict whether the user might want to take a break. 
The classifier can be refined using the user's actual logged breaks as explicit feedback on the quality of its predictions. 

\section{Formative Study}
\label{sec:study}

\rev{With a functional prototype of \artifactname in hand, we moved to Step 5 of the design science research process: \textit{evaluation}.
We were most interested in formative understanding of RQ2: the roles that researchers might imagine for mixed-initiative qualitative coding tools in mitigating the potential trauma they incur in their work.
Thus, our study took the form of interviews conducted with potential users of a tool like \artifactname, analyzed via a qualitative research process.}

\subsection{Methods}


\paragraph{Scenarios and procedure} In 1.5-hour semi-structured interviews, participants were asked to use \artifactname to annotate a synthetic dataset, within a provided scenario. The first scenario, called \textit{IPV case records}, prompted participants to imagine they were researchers receiving a dataset of client files from a clinic serving survivors of intimate partner violence (IPV).
These records included transcripts of conversations between survivors and their support workers.
Studying such records has enabled scholars to understand technology's role in IPV and how to better support survivors (cf. \cite{freed2019my, daffalla2023account, bellini2023paying, tseng2021digital, tseng2022care}). However, this data is known to contain details of survivors' abuse that may be traumatic for some analysts to read.
For the purposes of this study, we created three documents consisting of synthetic composites of case records. Each was hand-written by the first author based on their experience working in IPV.

The second scenario, called \textit{AOC's replies}, asked participants to imagine they were social media researchers studying direct responses to U.S. Congresswoman Alexandria Ocasio-Cortez (AOC) on the social media website Twitter (currently known as X.com). 
Among U.S. politicians, AOC is known to receive some of the highest amounts of hate, harassment, and violent threats online \cite{guerin2020public, casula2021we}.
Research on what Hua et al.~\cite{hua2020characterizing,hua2020towards} call ``\textit{adversarial interactions}'' with politicians online is necessary to understand the influence of social media on political discourse, and to protect the safety of politicians, their staff, and the democratic process. 
However, studying directed expressions of violence and abuse may be traumatic for an analyst---AOC, for example, is a woman of color known to receive racist and sexist threats \cite{casula2021we}.
For our study, we created a dataset of three documents, each of which contained the text of a tweet AOC posted in June 2023 alongside approximately 10 publicly-available direct responses.
Multimedia was omitted for the purposes of this study, which focused on qualitative coding of text.
Both scenarios are provided as supplementary material. 

Participants were asked to first code their synthetic dataset of choice using \artifactname, receiving suggested annotations from the tool (\figref{fig:tiqa_A}). 
Then, they were asked to turn on the traumatic exposure tracking (\figref{fig:tiqa_B}), and explore the reaction predictions and nudges.
Throughout, they were invited to ``\textit{think-aloud}'' \cite{lewis1982using} on their likes and dislikes, and reflect on the possibilities of the tool for use in their work.
\rev{The interview protocols are provided in Appendices \ref{app:protocol-ipv} and \ref{app:protocol-socmed}, and the data participants were asked to consider are in Appendices \ref{app:corpus-ipv} and \ref{app:corpus-socmed}.}

\paragraph{Recruitment and participants}
We recruited two categories of participants: (1) researchers who had experience working with firsthand accounts of intimate partner violence (IPV); and/or (2) researchers who had experience studying hate and harassment on social media.
These categories span two likely but meaningfully contrasting scenarios for \artifactname: the study of data drawn from intimate care encounters and the study of data posted publicly on the Internet.
\rev{Several participants had professional experience with both scenarios, and we did not want to assume which scenario they might prefer (e.g., some participants may have had undisclosed personal histories with IPV). Thus all participants were given the choice of which scenario to use in their interview.}

Participants were recruited via snowball sampling from email and professional listservs. 
In total, we interviewed 15 participants, of whom 9 chose the social media scenario, and 6 chose the IPV scenario.
\rev{\tabref{tab:participants} summarizes participants' relevant demographic and experiential characteristics.}
All had experience working on collaborative coding projects.
Most had tried commercial qualitative coding tools like NVivo, QDAminer, MaxQDA or Dedoose---but found them cumbersome, expensive, or insufficiently collaborative, and instead used customized spreadsheets. 
Three participants who worked in NLP and computational social science had used Prodigy, an ML-based annotation tool.\footnote{https://prodi.gy/}

\begin{table}[t]
\centering
\small
\resizebox{\columnwidth}{!}
{%
\begin{tabular}{@{}lllllll@{}}
\toprule
\textbf{ID} & \textbf{Age} & \textbf{Gender} & \textbf{Fields} & \textbf{Preferred qual coding tools} & \textbf{Selected scenario} \\ \midrule
P01 & 26 & Woman & HCI, social computing & \begin{tabular}[c]{@{}l@{}}Excel and Notion; disliked Dovetail\end{tabular} & AOC's replies \\ \midrule
P02 & 24 & Woman & HCI, social computing & Ad hoc Shiny apps & AOC's replies \\ \midrule
P03 & 23 & Woman & Security and privacy & \begin{tabular}[c]{@{}l@{}}Notion; disliked Dedoose\end{tabular} & IPV case records \\ \midrule
P04 & 26 & Man & Security and privacy & \begin{tabular}[c]{@{}l@{}}Google Sheets; disliked NVivo\end{tabular} & IPV case records \\ \midrule
P05 & 30 & Woman & Social work, policy & \begin{tabular}[c]{@{}l@{}}Spreadsheets, Post-It notes,\\verbal debriefs; disliked NVivo\end{tabular} & IPV case records \\ \midrule
P06 & 28 & Woman & Security and privacy & \begin{tabular}[c]{@{}l@{}}Spreadsheets; disliked NVivo\\ \& MaxQDA (too expensive)\end{tabular} & AOC's replies \\ \midrule
P07 & 27 & Woman & Security and privacy & \begin{tabular}[c]{@{}l@{}}Dedoose; disliked NVivo\end{tabular} & IPV case records \\ \midrule
P08 & 36 & Non-binary & HCI, social computing & Google Sheets & AOC's replies \\ \midrule
P09 & 29 & Woman & \begin{tabular}[c]{@{}l@{}}NLP, computational social science\end{tabular} & \begin{tabular}[c]{@{}l@{}}Google Sheets, Airtable, MaxQDA\end{tabular} & IPV case records \\ \midrule
P10 & 27 & Woman & \begin{tabular}[c]{@{}l@{}}NLP, computational social science\end{tabular} & Prodigy, Excel & AOC's replies \\ \midrule
P11 & 38 & Woman & HCI, social computing & \begin{tabular}[c]{@{}l@{}}Word, Excel, Google Sheets;\\disliked QDAminer (needed \\more collaborative features)\end{tabular} & AOC's replies \\ \midrule
P12 & 30 & Woman & Security and privacy & \begin{tabular}[c]{@{}l@{}}Excel, Google Sheets;\\ disliked NVivo (needed more\\ collaborative features)\end{tabular} & AOC's replies \\ \midrule
P13 & 31 & Woman & HCI, social computing & \begin{tabular}[c]{@{}l@{}}Excel, Google Sheets\\Post-It notes; disliked NVivo\end{tabular} & AOC's replies \\ \midrule
P14 & 34 & Woman & \begin{tabular}[c]{@{}l@{}}NLP, computational social science\end{tabular} & Prodigy, Potato & AOC's replies \\ \midrule
P15 & 25 & Woman & Security and privacy & Excel & IPV case records \\ \bottomrule
\end{tabular}%
}
\caption{Relevant demographic and experiential characteristics of study participants.}
\label{tab:participants}
\end{table}

\paragraph{Data collection and analysis}
The first author conducted all semi-structured interviews.
Each session was audio-recorded, with the participant's consent, and professionally transcribed.
The first author additionally took written notes and composed post-session memos for each participant.

Data for each participant (transcripts, notes, and memos) were analyzed by the first author using a thematic analysis approach adapted from Braun \& Clarke \cite{braun2019reflecting}.
We aimed to retain an open and inductive analytic process, in order to understand the fullness of participants' reactions to the provocation in the formative study. 
\rev{We also focused our analysis on answering RQ2: what \textit{roles} might mixed-initiative coding tools have in mitigating researchers' trauma?}
Major themes were developed by the first author and refined through multiple rounds of coding and discussion with the second and third authors.
The resulting themes are presented as findings in \secref{sec:findings}.

\paragraph{Ethics}
We took care to design study activities that, to the best of our ability, would not inadvertently induce greater trauma responses in our participants. 
All participant sessions were conducted by the first author, who has seven years of experience \rev{conducting research using trauma-informed principles (see \textbf{Positionality} in \secref{sec:intro}), and the interview guides, provocations, and scenarios were all constructed with these principles in mind.}
Participants were advised they could take a pause, stop the interview, or opt to not answer any question at any time.
\rev{Where necessary, we allowed interviews to take longer than the allotted one hour, to allow participants to fully share their prior traumatic experiences.}

We further took care to ensure we did not inadvertently compromise \rev{the privacy of any data subject or study participant}.
The IPV case record dataset was a fictional composite, written by the first author based on their 7 years of experience in IPV victim advocacy, and did not contain data from any real IPV survivors.
The social media dataset was all publicly posted. 
\rev{Both sets of composites are presented in the Appendix. The functional prototype itself used a locally hosted language model, and thus did not send any user inputs or fictional composite data to an external language model (see \secref{sec:design-implementation}).}
All study procedures were approved by our university's IRB.

\subsection{Formative Study Findings}

\label{sec:findings}

\begin{table}[]
\centering
\resizebox{\textwidth}{!}{%
\begin{tabular}{@{}|l|l|l|@{}}
\toprule
\textbf{Feature} & \textbf{Participant Feedback} & \rev{\textbf{Suggested Roles for TIQA}} \\ \midrule
\begin{tabular}[c]{@{}l@{}}ML-based workflow enabling a user \\ to train their personal model for \\ traumatic concepts\end{tabular} & \begin{tabular}[c]{@{}l@{}}Mixed-initiative tools can help users reflect\\ on their own coding practices---but only if \\ the biases inherent to language models \\ are addressed. (4.2.1)\end{tabular} & \rev{\begin{tabular}[c]{@{}l@{}}A self-reflective surface for \\ a coder's own analysis---NOT \\ an enforcer of efficient coding.\end{tabular}} \\ \midrule
\begin{tabular}[c]{@{}l@{}}Ability to initialize a model from a \\ trusted peer's model\end{tabular} & \begin{tabular}[c]{@{}l@{}}Affordances for peer-to-peer support are \\ valuable to a trauma-informed qualitative \\ coding workflow. (4.2.4)\end{tabular} & \rev{An initiator of collaboration.} \\ \midrule
\begin{tabular}[c]{@{}l@{}}Measurement of user's prior and \\ upcoming traumatic exposures, \\ based on their personal model \\ for traumatic concepts\end{tabular} & \begin{tabular}[c]{@{}l@{}}ML-assisted content warnings are a ``value-add'': \\ providing analysts new information on their own \\ traumatic exposures and reactions. However, \\ counts of potentially traumatic text snippets may \\ not be a valid measurement of a user's \\ exposure to distressing content. (4.2.2)\end{tabular} & \rev{\begin{tabular}[c]{@{}l@{}}A self-reflective surface for\\ an analyst's traumatic responses.\end{tabular}} \\ \midrule
\begin{tabular}[c]{@{}l@{}}Nudges towards self-care after \\ a certain exposure threshold\end{tabular} & \begin{tabular}[c]{@{}l@{}}Nudges can feel like workplace productivity \\ tools rather than tools for self-help. They can \\ help embed a culture of care in a team, especially \\ with junior analysts. However, they may only \\ be effective in the extreme, if used to lock an \\ analyst out of their workflow, and analysts \\ will likely bypass them anyway. (4.2.3)\end{tabular} & \rev{\begin{tabular}[c]{@{}l@{}}An initiator of a culture of care---\\ NOT an enforcer of strict breaks.\end{tabular}} \\ \midrule
\begin{tabular}[c]{@{}l@{}}\revtwo{(This feature was not in the} \\ \revtwo{implementation; several} \\ \revtwo{participants asked for it.)}\end{tabular} & \begin{tabular}[c]{@{}l@{}}TIQA could enable a head coder to allocate \\ documents between coders according to their \\ traumatic exposure---but only if privacy is \\ respected, both between coders and \\ between coders and their supervisors. (4.2.4)\end{tabular} & \rev{\begin{tabular}[c]{@{}l@{}}A trusted confidante and a fair \\ allocator of team responsibility---\\ with accountability to a human \\ supervisor.\end{tabular}} \\ \bottomrule
\end{tabular}%
}
\caption{Mapping of TIQA's features (left) to participants' feedback on their affordances (middle)\rev{, and what that feedback suggests about potential roles for machine assistance in mitigating researchers' potential trauma (right).}}
\label{tab:features-to-roles}
\end{table}

\rev{In line with prior literature, participants had strong opinions on how ML ought to be incorporated into their analytic practices. Still, they readily imagined individual and collaborative workflows using \artifactname to lessen the trauma possible in their research work. 
Our themes are summarized in \tabref{tab:features-to-roles} and detailed in this section.
In each subsection, we summarize participants' feedback on a particular \artifactname feature, and then describe what our analysis found that feedback suggests about the role \artifactname might play in mitigating researchers' trauma (RQ2).
The specific roles our analysis identified are bolded in the text.}


\subsubsection{Machine-assisted analysis should prioritize creative self-reflection over task efficiency}
\label{sec:findings-heavy}





Several participants initially had strong reactions against the idea of using ML assistance in their qualitative coding.
As in prior work (cf. \cite{jiang2021supporting, feuston2021putting, baumer2017comparing}), the very use of ML in exploratory analysis---however human-controlled and personalizable---seemed to sacrifice deep sensemaking for speed and scale.
As P14 said, ``\textit{I feel like I'm good at labeling things, and I don't want anything to come between me and the data.}''
For the study of heavy subjects, participants felt it was important to take one's time:

\begin{quote}
    ``\textit{If you go too fast, it feels like you're sanitizing someone's experience. I'm slower in this work. It should be slower when it's heavy.}'' (P12)
\end{quote}

Part of the problem, participants explained, was that automated code suggestions could push their analytic practice towards \textit{pattern-confirmation}, the phase of Nelson's framework where patterns identified in smaller datasets are tested against larger corpora.
Many had experience in collaborative coding to label large corpora, where inter-rater reliability (IRR) had been a desirable goal.
They observed code suggestions could possibly help analysts achieve IRR faster, by avoiding missing an instance where a code should have been applied, and more rapidly converging on the same conceptual framework as fellow raters.
While faster coding might have been desirable for some corpora and some research questions, several participants expressed concern that less-experienced analysts (e.g., student research assistants) might not do the necessary human validation, and instead rely entirely on the code suggestions.

Instead, participants wanted \artifactname to emphasize the creative and inductive phases of qualitative work.
Several said they would use suggestions for similar codes as a surface for reflection on their own definitions of concepts---were they accidentally coding examples that emphasized one aspect over another?
Participants were particularly interested in whether \artifactname could help them track how their own usage of certain terms differed from project to project. 

Interestingly, for this kind of use, participants were not concerned about how \rev{model evaluation}: how accurate the tool's suggestions seemed to be, \rev{how many annotations were needed to reach a given threshold of accuracy}, whether the model was overfitting, or how its predictions might be explained. 
Several said adding information like prediction uncertainty would simply clutter the interface.
Whether \artifactname actually found segments that reflected the code was beside the point: it was clear they could improve the system's suggestions by fine-tuning the models with more annotations, and in any case, they could simply ignore the tool's suggestions.
More important was whether the tool encouraged a user to reflect more deeply on their own coding patterns.

Ultimately, participants agreed creative exploration was their preferred use of ML assistance \textit{because of} its inherent limitations.
Statistical pattern recognition could provide an interesting interface for an analyst's reflection on their own work, but the model could not itself become the source of ground truth.
As P14, a machine learning expert, said:

\begin{quote}
    ``\textit{I just don't trust our models. Indeed, we need ML, because indeed, there's too many things to moderate, but if we leave it entirely to ML, we know it's going to make mistakes and be horribly biased.''} (P14)
\end{quote}

\rev{We viewed these findings as indicative of an important balance to strike in \artifactname's role in a coding workflow. Given the inherent limitations of statistical language models, a tool like \artifactname should err on the side of being a \textbf{self-reflective surface, rather than an enforcer}. It should encourage the creative exploration of qualitative coding---\citet{jiang2021supporting}'s ``\textit{serendipity}''---by encouraging a coder to reflect on their own practices, rather than emphasize task efficiency.} 

\subsubsection{Supporting analysts' self-management of traumatic exposure presents a ``\textit{value-add}'' for ML}
\label{sec:findings-trauma-track}

Despite concerns around the incursion of ML into qualitative analysis, when presented with how code embeddings and semantic search could be applied towards trauma mitigation, participants were nevertheless intrigued. One reflected:

\begin{quote}
``\textit{[Using ML to track traumatic exposure] is more palatable than in the analysis process, because these are not predictions I could have made. This is an actual value-add for ML that's not automating away the essential part of my work.''} (P06)
\end{quote}

The ``\textit{value-add}'' was especially clear in \artifactname's tracking of prior and upcoming traumatic exposure (\textbf{B3}, \figref{fig:tiqa_B} and \secref{sec:system-trauma}). 
Several said their current work practices already involved budgeting more time than usual for qualitative coding of heavy data, because they were not sure when they might need to decompress after reading traumatic stories. 
With data on upcoming exposures, participants said they could better plan their own self-care practices, e.g., scheduling a particularly trauma-heavy analysis for a day they could go for a run after work, or speak with a therapist.

Just as participants saw potential in \artifactname as a tool for self-reflection on their analytic practices (\secref{sec:findings-heavy}), they also saw \rev{an important potential role} in \artifactname's exposure tracking feature: as \textbf{a surface for self-reflection} on their own limits. As P07 said:

\begin{quote}
    ``\textit{I have this rough sense that certain codes will trigger some trauma more than others. I would want to have some concrete evidence from the software that could help me investigate that. I wonder if it can tell me something like whether this code is indeed giving me more trauma than that code.}'' (P07)
\end{quote}

In line with P07's wishes for ``\textit{concrete evidence}'' of suspected traumas, several participants mentioned wanting longitudinal data on their exposures, to supplement the exposure modeling done in \figref{fig:tiqa_B}.
Such longitudinal dashboards could help them reflect after the fact---e.g., exhausted after a long week, they could look back on their content exposures and see if they had encountered large amounts of potentially traumatic subjects.

While the utility of these affordances was clear, several participants were not sure how \artifactname could meaningfully measure exposure to achieve them.
\artifactname's proxies for prior and upcoming exposures included counts of exposure codes, but several participants had misgivings about whether such a discrete measurement could capture traumatization. P05, who has social work training, said:

\begin{quote}
    ``\textit{For me, I don't know if it's the quantity of incidences of a word or a concept, it's more to do with the impact on me, and how graphic it was. It could be just one incredibly graphic description of something horribly violent, and it might have more of an impact on me than repeated instances of the phrase assault.}'' (P05)
\end{quote}

Relatedly, participants stressed \artifactname's predictions for upcoming exposures had to be \rev{evaluated and} validated:
too many false negatives, and an analyst would be exposed to more traumatic content than they wanted to see; 
too many false positives, and an analyst would be inappropriately put on edge, waiting to expect traumatic content they would not encounter.
More complicated still, participants reflected that the only way to improve the accuracy of \artifactname's predictions would be to refine it, by providing more annotations for what an analyst considered traumatic and non-traumatic content---which would require the analyst to encounter more traumatic content.

\subsubsection{Analysts want help taking breaks, but nudges may be ineffective}
\label{sec:findings-trauma-nudge}
\artifactname's affordances for nudging users towards self-care prompted several participants to reflect on how badly they needed to take more breaks. 
Because their jobs relied on the study of intense and potentially emotionally disturbing topics, participants described a kind of valor in consuming as much traumatic content as they could find. As one said:

\begin{quote}
``\textit{There is a tendency to say `I can handle it, it’s fine.' But we also know there are things that weigh on us. So [the tool] might make me take more breaks than I would consciously impose on myself.}'' (P05)
\end{quote}

Overcoming an analyst's tendency to take on more than they could handle was seen as especially important on projects that included more junior analysts.
Several participants who had experience supervising coding teams said they tried their best to instill trauma-informed research practice in their collaborations, but often were not sure whether the advice was followed.
Building a tool with self-care nudges directly into the coding workflow would help create room for an analyst in training to safely learn how to observe these practices.

These nudges, however, require users to know enough about their own reactions to trauma to provide accurate feedback to ReactionPredicter.
Several said they were not sure all users could make the distinction between acceptable and unacceptable levels of traumatic exposure. As P01 asked, ``\textit{What's a healthy amount of trauma?}''
Others said it was difficult to recognize their own traumatic reactions in-the-moment:

\begin{quote}
    ``\textit{Maybe I haven't been introspective enough to notice a very clear pattern before I hit the point I need to take a break, 
    but...I think you don’t always notice which instance pushes you over the edge.
    }'' (P12)
\end{quote}

Even assuming a sufficiently ``\textit{introspective}'' user, perhaps with a dashboard of prior exposures (\ref{sec:findings-trauma-track}), participants said at the end of the day, nudges had a natural limit on their value.
First, to accurately predict a traumatic reaction, \artifactname would need to know not only the content of the text an analyst was reading, but also more on their environment and day-to-day life events.
As P08 said, ``\textit{maybe my tolerance one day is different from my tolerance another day.}''

More importantly, many participants said they would likely simply ignore nudges.
P12 said they would probably dismiss nudges the same way they ignored notifications from ScreenTime, an iOS feature that informs users how long they have spent looking at their phones. 
As an alternative, several participants asked whether the tool could \textit{force} an analyst to take a break, for example by displaying a solid overlay over the text, and providing links to curated resources for mental health and well-being. 
Several drew connections to productivity apps that block a user from visiting specific distracting websites for a specified amount of ``\textit{work time}''.
Still, participants reflected they would ultimately have read the entire dataset, and would likely find workarounds for such restrictions.

\rev{In short, participants' reactions to this feature outlined that a tool like \artifactname should strive to \textbf{initiate self-care practices without mandating them}. A mixed-initiative system can encourage analysts to take breaks, but must attend to how doing so risks the tool becoming an enforcer of workplace productivity metrics. 
}

\subsubsection{Measurements of traumatic exposure could be used to scaffold collaboration in trauma-informed ways.}
\label{sec:findings-trauma-collab}
Assuming a reasonable measurement framework, participants pointed out \artifactname could have great utility not only in an individual analyst's work, but also in collaborative coding projects.
Participants generally had positive reactions to \artifactname's affordances for initializing a code from a trusted peer's trained model.
Several also said they would use their own exposure and trauma threshold data to inform conversations with their peers about the effect of the work on their well-being, to catalyze a practice of peer support.
These peer-to-peer affordances could help build trust between teammates, participants said, and establish more empathetic and human connections that would enable deeper analyses.

\artifactname also seemed to have utility for supervisory contexts, to help distribute a corpus of data among a team of analysts.
Several participants asked for a cross-document dashboard using the upcoming exposure modeling to determine what traumatic content was yet to appear, so they could assign documents to analysts according to their particular sensitivities: e.g., a supervisor could avoid assigning a document with high amounts of content related to sexual assault to a team member who was a survivor of a similar experience.

While participants were excited about this potential, they also raised immediate privacy concerns. 
In both the peer-to-peer and supervisory contexts described above, participants wanted an analyst to have control over how much of their personal information \artifactname would store, disclose, and infer.
One participant, an expert on privacy, said that while measuring \textit{exposures} seemed possible in an aggregate statistic, measuring a team's \textit{reactions} to specific exposures seemed necessarily personal and privacy-compromising:

\begin{quote}
    ``\textit{You could try aggregating [exposure data], 
    like see overall how well your team is dealing,
    how often do they see [a traumatic subject like] racism.
    But you'd need to do that without directly attributing [a reaction] to anyone. I'm just concerned about someone building a profile on my ability to handle trauma.}'' (P08)
\end{quote}

By ``\textit{building a profile}'' on an analyst's traumatic reactions, \artifactname seemed to create a workplace monitoring technology that could threaten analysts' autonomy.
All participants said if they were supervising a research team, they would not \textit{want} to violate a junior team member's privacy by learning more about what subjects they had traumatic reactions to.
But to distribute workload fairly between a team, it seemed a supervisor would be \textit{required} to know this information---at the very least, to validate that \artifactname had not made a mistake.

\rev{In short, the role our participants saw for \artifactname in collaboration was complex: all at once, it should be a \textbf{trusted confidante} on an individual researcher's traumas and sensitivities; a \textbf{fair allocator} of traumatic exposures between team members; and an \textbf{initiator} of peer-to-peer collaboration, surfacing opportunities for one coder to initialize their model from another coder's.
All throughout, participants emphasized the tool should remain verifiable and accountable to a human (e.g., a supervisor should be able to validate the allocation proposed by the tool).}

\section{Discussion}
\label{sec:dis}


\rev{Our study suggests that by building \artifactname via a design process deliberately enacting TIC's principles (RQ1), we were able to create an artifact eliciting potentially beneficial roles for machine assistance in researchers' practices of self-care (RQ2).
The roles we identified, elaborated in \tabref{tab:features-to-roles}, are in alignment with the literature on mixed-initiative systems in qualitative coding, which emphasizes 
enabling the researcher to employ machine assistance but remain firmly in control (cf. \cite{jiang2021supporting, feuston2021putting, zerilli2019algorithmic}).
Our participants were clear about undesired roles for \artifactname: they did not want it to be an \textit{enforcer}, mandating efficient coding work or fully automating self-care.
They did, however, identify several desired roles for such a tool in both the individual and collaborative work of trauma mitigation: as \textit{a surface for self-reflection} on their own analytic insights and stress reactions, as an \textit{initiator} of support from their peers and supervisors, and as a \textit{fair allocator} of responsibilities between team members (\tabref{tab:features-to-roles}).

We view these roles as important future directions for how systems like \artifactname can impact research practices. 
For example, a system that helps a coder reflect on links or trends within their codes could deepen a coder's insight, thereby improving the quality of their analysis.
A system that reflects a coder's own stress reactions back to them may provide information they can use to develop their own coping mechanisms.
And a system might assist a supervising coder with specific tasks, like allocating documents to coders fairly or initiating support between peers, to help that coding team create the type of collaborative culture that can mitigate trauma.
Throughout, all of these systems must remain \textit{mixed-initiative} and, even further, \textit{accountable to human authority}: for example, a system for fair allocation of a coding workload must retain the ability for the supervising coder to check it for mistakes.
And they must \textit{never mandate} productivity or self-care in a coder's experience.
The human-machine systems that could fulfill these roles could go a long way to helping researchers mitigate their trauma.
}

\rev{ 
Realizing these benefits, however, will require conceptual and practical advancements to the trauma-informed design processes we undertook in this work.
In this section, we first unpack the conceptual tensions our work surfaced in how TIC can be translated to technology design, and argue for a refined lens on safety in software, towards \textbf{safety-as-enablement} (\ref{sec:dis-safety-collision}).
We then consider the \rev{practical} need to evaluate design processes with respect to trauma-informedness, and outline future work improving design processes in lieu of established measurement frameworks for trauma (\ref{sec:dis-eval}).
\revtwo{While both of these contributions emerged from the specific test case in the present study, software for qualitative analysis, we offer them as conceptual shifts for how \textit{any} sociotechnical system might adopt TIC, or otherwise become more trauma-informed.}
We close with limitations and further directions for future research (\ref{sec:dis-limit}).}


\subsection{Reframe trauma-informed computing from exposure reduction to \textit{safety-as-enablement}}
\label{sec:dis-safety-collision}


Existing literature on TIC has emphasized its principles should be considered not a checklist, but rather an orientation to computing (\citet{chen2022trauma}), a sensitizing concept (\citet{scott2023trauma}), or guidance for computing research (\citet{razi2024toward}).
Our enactment supports this emphasis---specifically, we found that the principles of \textit{safety, trust, enablement, peer support, collaboration, and intersectionality} were often too diffuse to be considered an interaction model in the style of \citet{beaudouin2000instrumental}, a set of interface design principles in the style of \citet{blair2008user}, or a set of interaction guidelines in the style of \citet{amershi2019guidelines}.
This does not mean TIC does not have utility for design and development. 
\rev{\citet{zheng2024s} showed trauma-informed principles could be identified \textit{retrospectively} in design processes, as inspiration for design goals, activities, and objectives.
\revtwo{\citet{ramjit2024navigating} applied trauma-informed care to computer security consultations by analyzing how an existing protocol could produce traumatic stress reactions, and redesigning the protocol towards the trauma-informed practices employed by mental health professionals.}
Our work here investigated how it can also be applied \textit{prospectively}, \revtwo{in the technical and design decision-making required to build software systems}. 

\begin{table}[]
\centering
\resizebox{\textwidth}{!}{%
\begin{tabular}{@{}|l|l|l|l|@{}}
\toprule
\textbf{TIC Principle} & \textbf{Rationale Favoring Safety} & \textbf{Rationale Deprioritizing Safety} & \rev{\textbf{Resolution via Safety-As-Enablement}} \\ \midrule
\textbf{Trust} & \begin{tabular}[c]{@{}l@{}}ML-powered content filters \\ can limit users' exposure to \\ traumatic content (safety+); \\ but introducing ML requires \\ improving users' trust in the \\ technology (trust-): "we know \\ it's going to make mistakes \\ and be horribly biased" (P14).\end{tabular} & \begin{tabular}[c]{@{}l@{}}Explainability tools offering \\ transparency into why TIQA filters \\ out some content could improve \\ trust (trust+), but also inadvertently \\ expose the user to potentially \\ traumatic content (safety-).\end{tabular} & \rev{\begin{tabular}[c]{@{}l@{}}A user could decide for themselves \\ whether to engage with the filters or \\ the explainability tools; therefore, \\ ML-powered content filters with \\ explainability can be considered \\ trauma-informed.\end{tabular}} \\ \midrule
\textbf{Enablement} & \begin{tabular}[c]{@{}l@{}}Automatic content filters limit \\ a user's exposure to traumatic \\ content (safety+), but remove \\ their ability to fully control \\ their technology experience \\ (enablement-).\end{tabular} & \begin{tabular}[c]{@{}l@{}}Customizable content filters can \\ increase user agency (enablement+), \\ but sourcing and labeling negative \\ examples can require traumatic \\ exposure (safety-).\end{tabular} & \rev{\begin{tabular}[c]{@{}l@{}}A user can decide for themselves\\ whether sourcing and labeling negative\\ examples is safe or unsafe. Thus,\\ customizable content filters can be\\ considered trauma-informed.\end{tabular}} \\ \midrule
\textbf{Collaboration} & \begin{tabular}[c]{@{}l@{}}Automatic filtering can limit \\ exposure to traumatic content \\ (safety+), but also limit a user's \\ ability to co-create their \\ experience (collaboration-).\end{tabular} & \begin{tabular}[c]{@{}l@{}}Inviting users to co-design their \\ content filters via annotation tools \\ (collaboration+) requires exposing \\ them to potentially traumatic \\ content (safety-).\end{tabular} & \rev{\begin{tabular}[c]{@{}l@{}}If safety is not limited exposure but\\ rather the ability to control the user\\ experience, the ability to co-create\\ one's own content filters can be \\ considered a trauma-informed feature.\end{tabular}} \\ \midrule
\textbf{Intersectionality} & \begin{tabular}[c]{@{}l@{}}ML-powered content filters \\ (safety+) may inadvertently \\ account for just one dimension \\ of a user's intersecting identities, \\ e.g., obscuring harmful content \\ related to a person's gender but \\ permitting content related to their \\ race (intersectionality-).\end{tabular} & \begin{tabular}[c]{@{}l@{}}A user can theoretically customize \\ their content filters to their unique \\ intersection of systems of oppression \\ (intersectionality+). But doing so \\ requires exposing them to potentially \\ traumatic content (safety-).\end{tabular} & \rev{\begin{tabular}[c]{@{}l@{}}If safety is not limited exposure but\\ rather the ability to control the user\\ experience, a content filter that is\\ customizable to a user's identities\\ can be considered trauma-informed.\end{tabular}} \\ \bottomrule
\textbf{Peer Support} & \begin{tabular}[c]{@{}l@{}}TIQA could help users manage \\ their own stress reactions \\ (safety+). But if users overrely \\ on the tool, they could withdraw \\ from the peer relationships that \\ previously filled that purpose \\ (peer support-).\end{tabular} & \begin{tabular}[c]{@{}l@{}}A tool could allocate a dataset \\ between coders to minimize each \\ coder's traumatic exposure (peer \\ support+). But doing so could \\ expose teammates' sensitivities \\ to each other or to their supervisors, \\ violating their privacy (safety-).\end{tabular} & \rev{Less clear how to resolve.} \\ \midrule
\end{tabular}%
}
\caption{Examples of how TIQA features viewed with slightly different rationales can trade safety for other TIC principles. \rev{In the \revtwo{rightmost} column, we show how reframing safety as enablement helps resolve the documented collision (see \ref{sec:dis-safety-collision}).}}
\label{tab:safety-rationales}
\end{table}

Our work surfaced a core problem in the possibility of prospectively applying TIC:} collisions between safety and each other principle that require an expansion of what we consider safety to mean.
The trauma literature understands \textit{safety} as both physical and psychological, encompassing both an internal sense of security and the absence of external harm \cite{chen2022trauma, scott2023trauma}.
In line with this definition, \artifactname's key affordances centered safety by reducing users' traumatic exposure, via features like semi-automated content warnings personalized to their specific sensitivities, and nudges towards self-care after personalized exposure thresholds (\tabref{tab:safety-rationales}).
Emphasizing safety in this way would seem vital to the core goal of the tool, to mitigate trauma, and how well it aligned with trauma-informed computing.
Yet, in analyzing how well \artifactname reflected TIC's six principles, we found these features created design rationales forcing tradeoffs between safety and every other TIC principle. 

Semi-automatic content warnings or nudges towards self-care embody the reduction of traumatic exposure, but also seem to hamper users' agency, violating the principle of \textit{enablement}.
This core tension between safety and agency was also echoed across other principles.
Take, for example, the basic premise of machine assistance in content warnings.
ML-powered warnings were seen to create problems of \textit{trust}, because it is well known that ML can ``\textit{make mistakes and be horribly biased}'' (P14).
Participants also perceived that less experienced researchers might overrely on machine assurances and forgo the work of personalizing the tool, limiting their \textit{collaboration} in creating their technology experience.
Such overreliance could also encourage users to withdraw from the peer relationships that previously supported their self-care, violating \textit{peer support}.
Lastly, with respect to \textit{intersectionality}, it is possible an ML-powered content warning system may account for just one of a user's intersecting identities, and fail to warn them of content related to another facet of their experience.

To complicate matters, slight variations in how a feature is viewed can lead an analyst to conclude that safety has been favored or deprioritized relative to other principles.
Take, for instance, the previous example of collisions between \textit{safety} and \textit{enablement}.
Rather than understanding ML-assisted content warnings to reduce \textit{enablement} in favor of \textit{safety}, an analyst could conclude that since the user has some control over \artifactname's features (e.g., turning them off, reducing notification frequency), the tool actually prioritizes \textit{enablement} at \textit{safety's} behest. 
This malleability is not localized to the tradeoff between safety and enablement: for every TIC principle, we found possible design rationales favoring or deprioritizing safety across \artifactname's initial affordances (\tabref{tab:safety-rationales}).
A tool could support users' needs for improved \textit{trust} in ML-powered content warnings by introducing tools for explainability and transparency---but for users to do the work of understanding the model through these methods, they would need to expose themselves to some form of content that could also be traumatic, harming safety.
This same exposure problem hampers potential design decisions that emphasize \textit{collaboration} and \textit{intersectionality}: co-creation of technology experience, for example by providing the tool with enough annotated examples to accurately reflect one's situated sensitivities, inherently requires traumatic exposure that could harm safety.
Finally, to encourage \textit{peer support} and deemphasize user-to-machine overreliance, a tool could build in collaborative workflows that allocate datasets between coders---but doing so would require attention to privacy between teammates, to avoid unwanted exposure of an individual researcher's traumas to their colleagues.

These tradeoffs indicate a need to shift the conceptual frame of \textit{safety}, to accommodate its complicated interaction with other principles.
We consider that safety in technologies for trauma mitigation is perhaps best understood less as the shielding or removal of harmful experiences, and more as the provision of tools enabling users to manage safety for themselves: a conceptual shift we call \textbf{\textit{safety-as-enablement}}.
Conceptualizing safety in this way 
draws our focus towards using traumatic exposure measurements as surfaces for self-reflection and collaboration, as our participants were inspired to do, and 
helps alleviate tensions between traumatic exposure and trauma-informedness (see \tabref{tab:safety-rationales}).
This widened aperture is also in concordance with the feminist approach to safety outlined in \citet{strohmayer2022safety}, which suggests embedding design friction into users' interactions with data-hungry platforms: specifically, the notion of a `\textit{trust pause}', during which users have a chance to critically reflect on whether to trust a system with their data.
A high-enablement interaction paradigm like this flouts typical design guidelines towards instantaneous reduction of burden via technology, but \citet{strohmayer2022safety} argue the short-term cost is to the longer-term benefit of maintaining users' trust in the system.

Importantly, safety defined as enablement may not be appropriate for all users: as prior work has described, in many situations, people may not want technologies to actually create extra work for them \cite{dombrowski2016social, tseng2020we}.
\rev{Framing safety as enablement also does not perfectly resolve every possible collision between safety and other TIC principles: as shown in \tabref{tab:safety-rationales}, exposure reduction was the rationale for deprioritizing safety for every principle except peer support.}
Still, as our work highlights, there may be unique benefits to encouraging enablement in technologies that scaffold potentially traumatic experiences: enablement is fundamental to trauma recovery (cf. \cite{chen2022trauma}).
We suggest that \textbf{safety-as-enablement} might ignite further design possibilities, both in qualitative coding software specifically and in trauma-informed design broadly, and encourage further interrogation of how TIC's principles might be adapted in this direction.

\subsection{Instead of measuring trauma, evaluate trauma-informedness of design processes}
\label{sec:dis-eval}
Compromising between principles might have been easier with a validated metric for the effect of a designed system on a user's well-being. 
However, neither academic literature nor best practices in trauma-informed care have converged on frameworks for such an evaluation.

Throughout the literature, there is a sense that trauma is a common human experience with serious negative repercussions---that at the same time defies precise measurement.
The range of potential human reactions to traumatic events is vast, and stress reactions are already known to be multifaceted.
Consider, for example, measuring traumatic stress by asking participants to complete psychometric tests for depression or anxiety before and after using \artifactname.
Such studies could provide a cross-sectional view of how well participants' experiences align with depression and anxiety symptomology (cf. \cite{das2020fast}), but risk flattening the experience of trauma into biological or behavioral signals of mental well-being that may not apply to every user. 
Allowing people to select their own signals is also not enough. As our participants pointed out, many people may not be aware enough of their own reactions to traumatic stress to choose, and indeed may want to use a self-reflective surface like \artifactname to investigate it: ``\textit{I think you don't always notice which instance pushes you over the edge}'' (P12).
There is also the broader issue of how to capture trauma's long-term effects.
Whether a system helps avoid re-traumatization requires assessment over an unknown time horizon: a person may have no reaction in the moment, but experience negative effects later, after reflecting on their experience.


The lack of validated measurement frameworks for trauma's effects may seem to stymie efforts to build trauma-informed technology at the root. 
However, we see instead an opportunity: rather than focusing on optimizing technological interventions with respect to to some measurement of user well-being that may never be settled, sociotechnical scholars can focus on \textbf{assessing the trauma-informedness of the \textit{design processes}} behind such tools, as a precursor to assessing the trauma-informedness of the resulting artifacts.
\rev{Our work here joins \citet{zheng2024s}: their retrospective analysis and our prospective inquiry provide instructive cases for the field on how trauma-informed principles can manifest in design. We encourage more scholarship demonstrating new approaches at this juncture.}

One outcome of a focus on the design process might mean that the creators of sociotechnical systems \textbf{document tradeoffs between principles} in the design of a system. 
In the design of \artifactname, for example, we elected to orient our design tradeoffs towards \textit{enablement}, as part of a broader commitment to \textit{intersectionality}:
our stance is that people know their personal and structural conditions best, and should be given the tools to craft their own experiences accordingly.
Documentation of these design tradeoffs would make apparent to stakeholders what the resulting artifact could reasonably be expected to afford its users.
Such a focus on process over outcome is not a solve for the lack of clarity on trauma's effects---as \citet{champine2019systems} highlight, it is a problem that there has been little investigation into whether trauma-informed approaches help organizations produce better outcomes.
However, in line with our previous suggestion to conceptualize safety as enablement, we consider that a transparent design process may at least help users craft their own trauma-responsive workflows around a technological artifact, towards practices of trauma mitigation that work for them.

\subsection{Limitations and Future Work}
\label{sec:dis-limit}

We conducted an exploratory qualitative study of how to mitigate traumatic exposure in qualitative coding software.
Our work is thus subject to the inherent limitations of formative qualitative design research: our research was a product of our participants' perspectives and our own interpretive lenses, and should not be interpreted as attempts to do broadly generalizable quantitative assessments of users' preferences.

In addition to the future work outlined in the rest of this Discussion, we see ample further work exploring the practical and theoretical issues invoked by \artifactname.
A longer-term study embedded in a collaborative coding team would further interrogate the temporal and interpersonal dynamics of traumatic exposure and mitigation in qualitative research.
\rev{
\artifactname is a design probe offering a technical intervention, and we see ample room for research on its corresponding social and organizational arrangements---for example, how might an annotation team organize its internal hierarchy to mitigate team members' trauma? 
In terms of user populations, we focused in our work on researchers, but a study with content moderators, journalists, or other potential users might also illuminate novel design goals.
There is also the possibility of applying safety-as-enablement and TIC in the landscape of research tools beyond qualitative coding: e.g., visualization and quantitative analysis of data may manifest different forms of researcher trauma, and present different opportunities for intervention.}

\rev{One particularly interesting direction is whether ML-backed text analysis systems like \artifactname can be implemented in a truly privacy-preserving way.
We avoided sending sensitive data to an external API in our study by implementing a small and locally hosted model.
We argue that systems with trauma-informed goals should prioritize this type of privacy preservation instead of using commercial LMs, where user inputs may end up as training data without their consent.
Balancing this important goal against the convenience of an external LM is an open question we reserve for future work.}

\revtwo{Another interesting direction is how to balance safety-as-enablement against trauma mitigation in machine learning systems like TIQA that require substantial user-level feedback. As our participants pointed out, refining a model like the sentence embeddings suggested in TIQA requires a user to provide additional annotations, and thus potentially expose themselves to even more traumatic content. Yet, providing feedback is necessary to evaluate the system's efficacy at identifying traumatic content---and especially to ensure two users with two different sets of starting embeddings and sensitivities can experience the same level of baseline efficacy. Striking a balance between personalized evaluation and TIC is, in our view, a key area of future work.}
Lastly, our work has focused on mitigating traumatic exposure for researchers, but there is ample future work to be done in parallel on how to mitigate research-related traumas experienced by research participants.

\section{Conclusion}

We explored how the tools and workflows of qualitative analysis might be redesigned to mitigate researchers' traumatic exposure.
In our Research-through-Design study, we first used the principles of \textit{trauma-informed computing} to redesign qualitative coding tools towards two design goals: (1) personalizable content warnings; and (2) self-management of traumatic exposure. 
We then used our resulting prototype, \artifactname, in an elicitation study with 15 researchers with experience qualitatively coding dark and disturbing content.
From our participants' reactions to the provocation, we synthesize considerations for the space of sociotechnical design around trauma mitigation in qualitative coding, and reflect on TIC's utility in the practice of software development.
Our work extends CSCW's literature on knowledge infrastructure towards a new dimension of concern: how scientific workflows can account not only for imperatives towards scale, but also to protect researchers' well-being.
\begin{acks}

This work was funded by the National Science Foundation via grant \#1916096.
ET was additionally supported by a Microsoft Research PhD Fellowship.
Thank you to our reviewers for excellent recommendations on how this work could link multiple literatures.
Tremendous gratitude also to Harini Suresh, Jonathan Zong, Justine Zhang, Michaelanne Thomas, Anne Jonas, Mark Ackerman, Ian Ren\'e Solano-Kamaiko, Sharon Heung, Marianne Aubin Le Qu\'er\'e, and the Social Media Collective at Microsoft Research, for uplifting discussions as this work developed.
Special thanks to Nancy Salem, Merrill Fabry, Zo{\"e} Glatt, and Richard Stupart, for generously sharing references on flesh witnessing and how journalism handles the problem of traumatic exposure.
Last but certainly not least, thank you to our participants for devoting their time, energy, and expertise to our study, and to our broader research community for their interest in and support of this effort.

\end{acks}

\bibliographystyle{ACM-Reference-Format}
\bibliography{references}

\appendix

\section{Interview Protocol: IPV Scenario}
\label{app:protocol-ipv}

\subsection{Introduction}

Hello and thank you for joining today! Is now still a good time for a 75-minute interview?

(If consent form is not yet on file) I don’t think I received a consent form from you, so let’s take a moment now to go over it. Let me know if you have any questions. 

(If consent is given) Thank you for that. So before we start the session, just a couple of demographic questions for study purposes. 
\begin{enumerate}
    \item What is your age?
    \item What is your gender?
    \item What is your current professional role?
    \item How long have you been doing qualitative research generally? 
    \item What qualitative coding tools and software do you use?
    \item How long have you been doing research with intimate partner or gender-based violence in particular?

\end{enumerate}

Let’s get started.
(If on Zoom, share researcher’s screen)
(If in-person, participant will sit at a laptop where the software is already up) 

Let me quickly introduce our study. We’re interested in designing tools to support researchers like yourself, who study text datasets that may be sensitive or difficult. For example, today we’ll be focusing on a synthetic dataset of transcripts from a clinic serving survivors of intimate partner violence (IPV). Our tool envisions that the data have not necessarily been collected by you—you have received them in a secondary release. Another entity, the data steward, is officially responsible for the data, and has given you this tool to use to work with it.

In our time today, I’ll show you some very early designs for particular features of this tool. I’ll ask you to play with it as though you were doing research on this synthetic data, and ask you specific questions. Throughout, please share if you have impressions about how the various features might be useful to you, or how they might create barriers to your work.

Do you have any questions about this? 

(Start recording if consented)

\subsection{Document Screen}

What you see here is how our tool would support analyzing one document. As you can see, this transcript has been de-identified. You see `Client' instead of the client’s name.

\begin{enumerate}
    \item Are there specific kinds of research you would want to do with this data?
    \item (If none) Let’s imagine you've been given this data to study, broadly, the experience of survivors of technology abuse. You will analyze this data using this tool. (Short description of specific affordances.)
    \item (If they have some, pick the one most amenable to qualitative coding.)
\end{enumerate}

You can highlight a section of text and attach codes. Please go ahead and do so. (Participant makes first annotations.)

(Tool processes annotation and highlights snippets in the rest of the document that are similar)

Now, you see the tool has made automatic suggestions for other sections in the text where your code might also be applicable.

\begin{enumerate}
    \item What are your initial impressions of this feature?
    \item How much does this workflow align with and differ from how you usually explore a text document?
    \item What do you make of the automatic suggestions, specifically?
    \begin{itemize}
        \item The suggestions are generated with a machine learning model. How would you like it to explain its suggestions?
    \end{itemize}
    \item Currently, you’re able to refine the model by accepting its suggestions. Does this workflow make sense to you? 
    \item What other kinds of feedback would you like to provide to improve the model?
    \item What would you change about this feature?
\end{enumerate}

\subsection{Exposure Tracker}

(Some questions to guide a discussion of their own practices for mitigating traumatic exposure)

\begin{enumerate}
    \item In your work doing qualitative analysis, using (insert the tools and software they mentioned at the start), do you take any steps for practicing self care / ensuring your own well-being? 
    \item How do you manage your exposure to traumatic/upsetting content right now?
    \item How did you learn this? Was there a particular training, etc?
    \item Have you had an experience where you were emotionally drained or otherwise had to manage how doing the research affects you?
\end{enumerate}

Now, I want to draw your attention to the ``Track Exposure'' button here with each code. This tool tracks how much of this concept you have been exposed to in the course of this data analysis. 

(Click code, show the red panes)

\begin{enumerate}
    \item What are your initial impressions of the exposure tracking tool?
    \item Does the exposure tracking metaphor make sense to you for codes like this? 
    \item The tool currently tracks exposure as the function of: (a) time spent on this task; (b) amount of annotations reflecting this concept; and (c) how difficult this concept was for you to read. Given your own experience of reading difficult content for research — does measuring exposure to traumatic content through these variables make sense to you? Why or why not?
    \item How else do you think the tool should measure your exposure to traumatic content?

(If you click ``I'm taking a break'', the tool learns the circumstances under which you want to take a break, and suggests that you do so in future similar scenarios.)

    \item How would you imagine this feature changing your current practice of qualitative analysis?

Please continue annotating the document. (Participant adds more annotations. The tool lights up where subsequent sections of the document may also contain the tracked code). The tool is giving you suggestions for when your tracked code will appear in the rest of the document. 

    \item How would you imagine this feature changing your current practice of qualitative analysis?

\end{enumerate}

\subsection{Wrap-up}

Thanks for your time today! Just two high-level questions to wrap up. 

\begin{enumerate}
    \item What are your overall thoughts on the merits and drawbacks of this tool?
    \item Does this tool change the way you think about studying this kind of data?
    \item What other features, kinds of data or use cases do you think these ideas would be useful for?
\end{enumerate}
\newpage

\section{Corpus: IPV case records}
\label{app:corpus-ipv}

\subsection{Case Record \#1}

Client: Hi, thank you for seeing me.

Consultant-01: Thank you for being here! What brings you in today?

Client: So, about a year ago, I started to realize something was not right. Actually, let me start before that. My ex and I were not in a good place. Let me just - let me tell you - let me pause for a moment and explain. You might have to be patient with me.

Consultant-01: Take all the time you need.

Client: (Crying) I'm sorry, I'm really sorry. Talking about my assaulter is just really, really difficult.

Consultant-01: It's no problem. You can stop or pause anytime you would like.

Client: I want to tell you all about it, it's just that since he started all this spying and...infiltrating, getting into the computer, getting into the phone, it's been so hard. I think he's hacked me. I think he's hacked the kids.

Consultant-01: Take all the time you need.

Client: Thank you. (Takes a deep breath). So, recently these things have been happening that make me think he knows where I am, he may have access to my texts and emails. Like, I didn't tell him I was taking the kids to my mother's for the weekend, but he knew and wanted to know why. It's making me worried, because I talk to my lawyer that way too, and I'm thinking, what if he knows what I'm saying about him to my lawyer?

Consultant-01: We can certainly look into that for you. Can you tell me how many text or email accounts you have? We can go through them one by one to see whether something's going on.

Client: Okay, thank you. I have a Gmail, I think. There's also an old Yahoo account, but I'm not sure whether that's the one that's on my phone. And I have one phone, it's a Samsung phone, I'm on a Mint Mobile plan. Oh, that's the thing, I don't know if my assaulter has access to my pictures too, like maybe that's how he knows where I'm going?"

\subsection{Case Record \#2}

Case 0498, appointment held on June 19th, 2023.

Client-0498: Hello?

Consultant-01: Hello?

Client-0498: Can you hear me?

Consultant-01: Hello?

Client-0498: Can you hear me?

Consultant-01: Yes, I can hear you.

Client-0498: Thank goodness. I thought it might've been another phone I have to replace.

Consultant-01: No, no, I can hear you just fine. We're connected now. Thank you for your patience. What brings you in today?

Client-0498: Well, I've had to keep replacing my phones. I have a situation with my ex-husband. I don't know how he keeps getting to the new phones, but something's not right. He has a new girlfriend who's half his age, and maybe it's her, I heard she works in IT.

Consultant-01: I'm sorry you're going through that situation. Can you tell me more about what feels not right, why you keep having to replace your phones?

Client-0498: I take a photo and it disappears. I take a video and it disappears. Maybe they know I have the proof and then they have her hack in, I don't know. I don't know how any of this stuff works, I'm not a tech-y person.

Consultant-01: We can certainly try to help you with that. Can you tell me what kinds of phones you have?

\subsection{Case Record \#3}

Case 0425, appointment held on June 23rd, 2023.

Participant-0425: It's been full-on since then. He just keeps coming back and coming back and coming back. I don't have any clue how he's getting in, but I know I'm not the one changing my photos, it has to be him.

Consultant-01: I understand that must be frustrating.

Participant-0498: Like you wouldn't believe. I had a Google. He got in. I had a Yahoo. He got in. I even had my son make me another Google, still, he knows exactly what I've said to my lawyer, exactly where I'll be.

Consultant-01: We can definitely try to help you with that. Can you help me understand, how many email accounts do you have?

\newpage

\section{Interview Protocol: Social Media Scenario}
\label{app:protocol-socmed}

\subsection{Introduction}

Hello and thank you for joining today! Is now still a good time for a 75-minute interview?

(If consent form is not yet on file) I don’t think I received a consent form from you, so let’s take a moment now to go over it. Let me know if you have any questions. 

(If consent is given) Thank you for that. So before we start the session, just a couple of demographic questions for study purposes. 
\begin{enumerate}
    \item If you'd be comfortable sharing with me---what is your age and gender?
    \item What is your current professional role?
    \item How long have you been doing qualitative research generally? 
    \item What qualitative coding tools and software do you use?
    \item How long have you been doing research with social media in particular?
\end{enumerate}

Let’s get started.
(If on Zoom, share researcher’s screen)
(If in-person, participant will sit at a laptop where the software is already up) 

Let me quickly introduce our study. We’re interested in designing tools to support researchers like yourself, who study text datasets that may be sensitive or difficult. 

For example, today we’ll be focusing on a synthetic dataset representing an archive of hate speech on social media. Our tool envisions that the data have not necessarily been collected by you—you have received them in a secondary release. Another entity, the data steward, is officially responsible for the data, and has given you this tool to use to work with it.

In our time today, I'll show you some very early designs for particular features of this tool. I’ll ask you to play with it as though you were doing research on this synthetic data, and ask you specific questions. Throughout, please share if you have impressions about how the various features might be useful to you, or how they might create barriers to your work.

Do you have any questions about this? 

(Start recording if consented)

\subsection{Document Screen}

What you see here is how our tool would support annotating one document containing this Twitter thread.

\begin{enumerate}
    \item Are there specific kinds of research you would want to do with this data?
    \item (If none) Let’s imagine you’ve been given this data to study, broadly, hate speech towards political figures. You will code this data using this annotation tool. (Short description of specific affordances.)
    \item (If they have some, pick the one most amenable to qualitative coding.)
\end{enumerate}

You can highlight a section of text and attach codes. Please go ahead and do so. (Participant makes first annotations.)

(Tool processes annotation and highlights snippets in the rest of the document that are similar)

Now, you see the tool has made automatic suggestions for other sections in the text where your code might also be applicable.

\begin{enumerate}
    \item What are your initial impressions of this feature?
    \item How much does this workflow align with and differ from how you usually explore a text document?
    \item What do you make of the automatic suggestions, specifically?
    \begin{itemize}
        \item The suggestions are generated with a machine learning model. How would you like it to explain its suggestions?
    \end{itemize}
    \item Currently, you’re able to refine the model by accepting its suggestions. Does this workflow make sense to you? 
    \item What other kinds of feedback would you like to provide to improve the model?
    \item What would you change about this feature?
\end{enumerate}

\subsection{Exposure Tracker}

(Some questions to guide a discussion of their own practices for mitigating traumatic exposure)

\begin{enumerate}
    \item In your work doing qualitative analysis, using (insert the tools and software they mentioned at the start), do you take any steps for practicing self care / ensuring your own well-being? 
    \item How do you manage your exposure to traumatic/upsetting content right now?
    \item How did you learn this? Was there a particular training, etc?
    \item Have you had an experience where you were emotionally drained or otherwise had to manage how doing the research affects you?
\end{enumerate}

Now, I want to draw your attention to the ``Track Exposure'' button here with each code. This tool tracks how much of this concept you have been exposed to in the course of this data analysis. 

(Click code, show the red panes)

\begin{enumerate}
    \item What are your initial impressions of the exposure tracking tool?
    \item Does the exposure tracking metaphor make sense to you for codes like this? 
    \item The tool currently tracks exposure as the function of: (a) time spent on this task; (b) amount of annotations reflecting this concept; and (c) how difficult this concept was for you to read. Given your own experience of reading difficult content for research — does measuring exposure to traumatic content through these variables make sense to you? Why or why not?
    \item How else do you think the tool should measure your exposure to traumatic content?

(If you click ``I'm taking a break'', the tool learns the circumstances under which you want to take a break, and suggests that you do so in future similar scenarios.)

    \item How would you imagine this feature changing your current practice of qualitative analysis?

Please continue annotating the document. (Participant adds more annotations. The tool lights up where subsequent sections of the document may also contain the tracked code). The tool is giving you suggestions for when your tracked code will appear in the rest of the document. 

    \item How would you imagine this feature changing your current practice of qualitative analysis?

\end{enumerate}

\subsection{Wrap-up}

Thanks for your time today! Just two high-level questions to wrap up. 

\begin{enumerate}
    \item What are your overall thoughts on the merits and drawbacks of this tool?
    \item Does this tool change the way you think about studying this kind of data?
    \item What other features, kinds of data or use cases do you think these ideas would be useful for?
\end{enumerate}
\newpage

\section{Corpus: AOC's replies}
\label{app:corpus-socmed}

\subsection{Document \#1: 06-08-2023 January 6th}

Snapshotted from https://twitter.com/RepAOC/status/1656740505474985984?s=20 on June 8th, 2023.

Tweet: Rep. Alexandria Ocasio-Cortez @RepAOC: January 6th was just a dress rehearsal. For as bad as January 6th is, I believe that former President Trump would not have qualms about going further. Without a shadow of a doubt.

Replies:

Toxic Something Podcast @KeithBurgin: What you believe is of little consequence.

Progressive Angel @AngelWaterton90: Exactly. Make AOC Speaker of the House to promote a strong progressive Democratic messaging. AOC is the best communicator and the best online representative the Democrats have in the House of Representatives. Democrats shouldn't hide her, they should feature her.

pisic @pisic: Relax baby

TheKellyJaye @KellyJaye: You. Are. A. Disgrace.

Darby @NewPresident696: This woman represents a platform for feminazi's.

Oliviathegrey @oliviathegrey: He flat out, said that he would acknowledge it if it was fair and true. Nobody's gonna except lies and rigged voting. Get real lady. People work hard to become President of the United States. It's not something to take lightly!

Brother Prodigal @BrotherProdigal: This from a bartender photoed in a bar with her legs held wide open? In what looks like her underwear while sporting her now famous 'Crazy' grin? The things that make you say 'Hmmmmmmmmmmmmmm!'

Brian McNicoll @McNicollb: Oh absolutely. Now could you be a dear and grab me a scotch neat.

Timothy Vickery @TVickery19: You're a race traitor to yourself.

\subsection{Document \#2: 06-07-2023 Black disenfranchisement}
Snapshotted from https://twitter.com/RepAOC/status/1666490992210624513 on June 8th, 2023.

Tweet: Rep. Alexandria Ocasio-Cortez @RepAOC: Republicans are so afraid of democracy that they want to disenfranchise predominantly Black voters who have been disenfranchised for as far back as when Black people were enslaved in the USA. This has nothing to do with election integrity. This is about racial control.

Replies:

The Redheaded libertarian @TRHLofficial: Props on wearing Jeffrey Dahmer's glasses when you talk about hurting black men.

Justin T. Haskins @JustinTHaskins: You're a disgusting person. These are disgusting lies. Asking people to prove their identity when voting so we know they aren't voting numerous times or that they really live in the district is not racist. You want voting to be less strict than signing kids up for little league.

Nem Nova @Nem\_Nova: Screaming banshee says what?

Danny's twit @DannyMckeighen: Dang, dirt has a higher IQ than AOC

Quincy Lee Stephen Bingham @quincylsb: This is not the parody account right?

Mustlovecats @Norma\_Jea\_\_\_: Are you gonna cry about it?

Coach Corey Wayne @CoachCoreyWayne: Low IQ take of the day.

Brandoniduni @BrandonElect: Always a racist!

Truth Refugee @truthrefugee: You are a liar and a racist.

Professor Terguson @WeaselArmy: Please just move to Cuba.

david ds @Daviddss96: Woman it's only logical to require some sort of proof that you are who you say you are when it comes to voting

Goodbyecruel.wrld @fluxyjbarn: I mean if you wanted to help, I feel like democrats had like 50 years to do something. Plus the first 2 years of biden yall had full control to literally do whatever... am I missing something? I mean fix the education system in low income areas at least right?"

\subsection{Document \#3: 06-22-2023 DeSantis}

Snapshotted from https://twitter.com/RepAOC/status/1671994213628534784 on June 23, 2023

Tweet: Rep. Alexandria Ocasio-Cortez @RepAOC: Doesn't matter if you're red or blue, woke or asleep. When people speak at the ballot box their will is honored...but Ron DeSantis says, Floridians, I know what's better for you than you do.' After FL voters raised the min wage \& expanded voting, DeSantis blocked the new laws.

Replies:

No Dem Left Behind @NoDemLeftBehind: Republican officials hate democracy.

Shane Hatch @hatch\_shane: So- in 1950's a cheese burger costed about .20 cents.  Now it costs \$5.  What changed? Inflation creates the need to raise wages, yet raising wages creates inflation.  There is no economic way to have a 'living' minimum wage.  Min wage is for teens. It's a way to learn and get paid.  It's not and can never be a living wage.

Adam Parker @akparker: Seems to me, DeSantis' position is that citizens know their needs better than the government does. I think FL citizens agree, since they used the ballot box to put him there.

Nancy Hitchcock @HitchcockDries: You should go back to tending bar!!

David @marshalllaw13: Cortez you are a waste of taxpayer time and money

@Gary\_Roc @gary\_roc: The people wanted Amazon to build a headquarters in New York, so they can provide jobs and you 'knew what's better for them, then they do'

Elle Lorraine @elle\_lorraine: Thank you for caring about Florida. It is terrible here. He has taken so many things away from us. He cut millions in his budget that we need in the Tampa Bay area. He is spiteful. He's sneaky. We're frustrated.

Mr. White @MrWhiteMAGA: quit being fake worried about DeSantis...you are afraid of Trump only

RVIVR @RVIVRdotcom: What percentage of the Florida vote did DeSantis get, hon?

M @TheMikeMind: What does this have to do with NY? Stay in your lane.

Dark Star @DarkStarMach10: You are a pretty communist.

peter Georgiou @realGeorgiou: Negroni please. You need one to actually listen to AOC.  LMAO.

Andrea E @AAC0519: Because Democrats cannot be trusted to honestly count ballots.  It's actually very simple, @AOC

Jac Jax @Starfoxy32: How about take care of your constitutes baby girl we are fine here in FL that's why people from your state are moving here in the 100s of thousands

Maxgull @maxgull: Refresh my memory @RepAOC, aren't you singularly responsible for keeping Amazon out of NYC, hundreds of good paying jobs and million\$ in revenue lost? Yeah, that was you. Sit down.

Pops62 @Pops5662: What we called 'almost pretty' in college.

Tomas @T0MAS1957: Every politician purports to know better what people need than people know themselves.  That is what they do - including you and your cult leader.

Why am Im I back on here @Beasl10Leigh: Didn't she prevent Amazon from opening a warehouse in her district which would of gave her people jobs, this is the same lady, right. SMH

\newpage


\end{document}